\documentclass[fleqn,usenatbib]{mnras}

\usepackage{newtxtext,newtxmath}
\usepackage{multirow} 

\usepackage[T1]{fontenc}
\usepackage{amsmath}
\newcommand{\numberthis}{\addtocounter{equation}{1}\tag{\theequation}}
\DeclareRobustCommand{\VAN}[3]{#2}
\let\VANthebibliography\thebibliography
\def\thebibliography{\DeclareRobustCommand{\VAN}[3]{##3}\VANthebibliography}

\usepackage{graphicx}	
\usepackage{amsmath}	
\renewcommand{\vec}[1]{\boldsymbol{#1}}

\title[Linear rotating magnetic tides]{Tidal dissipation in magnetised, rotating stars and planets: linear calculations exploring various magnetic field configurations}

\author[S.~Chu, Z.~Guo, A.~Astoul, A.J.~Barker, R.~Hollerbach]{
Shijun Chu,$^{1}$\thanks{E-mail: S.Chu@leeds.ac.uk (SC)}
Zhao Guo,$^{1}$
Aur\'{e}lie Astoul,$^{1,2}$ 
Adrian J. Barker$^{1}$\thanks{E-mail: A.J.Barker@leeds.ac.uk (AJB)} \&
Rainer Hollerbach$^{1}$
\\
$^{1}$ School of Mathematics, University of Leeds, Leeds LS2 9JT, UK \\
$^{2}$ Institut de Recherche en Astrophysique et Plan\'{e}tologie (IRAP), Universit\'{e} de Toulouse, CNRS, Toulouse, France
}

\date{Accepted Nov 6 666. Received Nov 6 666.; in original form Nov 6 666.}

\pubyear{\the\year{}}

\begin{document}
\label{firstpage}
\pagerange{\pageref{firstpage}--\pageref{lastpage}}
\maketitle

\begin{abstract}
We study tidal flows in the convective envelopes of rotating, magnetised fluid bodies, such as low-mass stars and giant planets. In well-mixed convective regions, (magneto-)inertial waves are linearly excited by tidal forcing, and their dissipation can dominantly drive spin and orbital evolution in many close star-planet and binary star systems. We perform linear magnetohydrodynamic calculations of wavelike tides in spherical-shell geometry of a tidally-forced, rotating, incompressible, viscous and non-ideal magnetised fluid. Our calculations consider the widest range of magnetic field configurations to date (including both aligned and misaligned dipole fields, free-decay dipole and quadrupole fields, azimuthal ``Malkus fields'' and mixed poloidal-toroidal ``Prendergast fields'') to analyse the effects of magnetic fields on the wavelike response and dissipation. We find that the tidal response at a given frequency depends strongly on both magnetic field strength and geometry. Magnetic fields with strong poloidal components modify the flow more efficiently and introduce high-frequency Alfvénic resonances associated with weakly damped eigenmodes. When an enhanced (turbulent) viscosity is adopted, we find that viscous dissipation remains comparable to Ohmic dissipation for strong fields, in contrast to previous studies in which Ohmic dissipation was argued to dominate. We also explore the variation in magnetic effects as the shell thickness, magnetic Prandtl and Ekman numbers are varied. Finally, the frequency-averaged tidal power is found to be largely insensitive to the magnetic field in most cases, though significant deviations are found for free-decay fields. Our results have important implications for the tidal evolution of magnetised, rotating stars and planets.
\end{abstract}

\begin{keywords}
planet-star interactions -- stars: low-mass -- planets and satellites: gaseous planets -- (magnetohydrodynamics) MHD -- waves -- stars: magnetic field
\end{keywords}



\section{Introduction}

Tidal flows are excited inside stars and planets by gravitational forcing from their orbiting companions in star-planet and binary (and other multiple) star systems, and their dissipation causes exchanges of energy and angular momentum between spins and orbits, thereby producing important spin-orbit evolution \citep[e.g.][]{ogilvie2014tidal,Mathis2019,Barker2026}. In well-mixed (approximately neutrally-stratified) regions of rotating fluid bodies, tidal forcing often excites inertial waves. These waves are restored by Coriolis forces and have frequency magnitudes $|\omega|\leq 2|\Omega|$, where $\Omega$ is the rotation rate of a star or planet (which may depend on location). The dissipation of these waves in particular is believed to explain the circularisation periods of solar-type binary stars \citep[e.g.][]{Barker2022,Sethi2026,Dewberry2026} and to shape the eccentricity distribution of hot Jupiters \citep[e.g.][]{Nils2023,Lazovik2024}. They can also be important for the orbital evolution of close-in planets orbiting young, rapidly rotating stars \citep[e.g.][]{BM2016,gallet2017tidal,B2020,Ahuir2021,Lazovik2021}, as well as for the evolution of stellar and planetary spin-orbit angles \citep[obliquities, e.g.][]{Lai2012,B2016,LO2017,Damiani2018,Spalding2022,Nils2025,Spejcher2025}.

Stars and planets are believed to be magnetised objects \citep[e.g.][]{Mestel1999,DL2009,Jones2011,schubert2011planetary}, as indicated by spectropolarimetric and Zeeman-Doppler observations of stellar surface fields \citep{Koc06}, asteroseismic evidence for strong magnetic fields in red-giant cores \citep[e.g.][]{Fuller2015,Stello2016,Li2022}, and in situ spacecraft measurements of magnetic fields in Solar System planets \citep[e.g.][]{S2003}. The geometry of large-scale magnetic fields in solar-like stars has been inferred to be rather complex, with mixtures of poloidal/toroidal and axisymmetric/non-axisymmetric components, which can also vary with dynamo cycles and also on longer timelines associated with stellar evolution  \citep[e.g.][]{KP2017,BL2025}. While (usually young) fast rotators are often found to have a dominant (primarily axisymmetric) large-scale toroidal magnetic field \citep[this is also the case for early or late M dwarfs, e.g.][]{DJ2008,MD2010}, (older) slower rotators are typically found to have a primarily poloidal magnetic field structure \citep[e.g.][]{PD2008}, including dominant dipolar fields with kG strengths for some mid-M dwarfs \citep[][]{MD2008}. Strong quadrupolar and octupolar fields, often tilted with respect to the rotation axis, have also been detected in some young stars \citep[as in T-Tauri, e.g.][]{DJ2007,GM2008,GD2012} which can host close exoplanets \citep[e.g. as in the Kepler-78 system,][]{MD2016}.

This evidence indicates that it is essential to understand the effects of magnetic fields on their tidal responses \citep[e.g.][]{Astoul2019}. Magnetic fields can have a number of effects, including modifying the shape and tidal deformation of a body (if they are particularly strong), and in modifying tidal flows themselves. They are likely to be particularly important in modifying the tidal wavelike response in convective envelopes, where they modify the waves to become magneto-inertial \citep[or MC, for ``Magneto-Coriolis'', waves, e.g.][]{finlay2008course} and introduce an additional dissipation channel via Ohmic diffusion. Despite the likely importance of magnetic fields and their prevalence in astrophysical objects, their effects on tidal flows have been very minimally explored to date. 

\citet{lin2018tidal} and \citet{Wei2018} performed pioneering linear numerical magnetohydrodynamic (MHD) calculations to explore the effects of an imposed rotationally-aligned dipole field \citep[using the techniques to analyse free Alfvén modes developed in][]{RR2003}, or a constant vertical field, on the tidal response in incompressible rotating spherical shell models of the convective envelopes of stars and planets. The effects of other, more complex, and potentially more realistic field configurations have not yet been explored. In addition, their models were primarily restricted to deep shells, with an inner/outer radius ratio $\alpha=0.5$, although \citet{lin2018tidal} explored the frequency-averaged dissipation as a function of $\alpha$ for the case of an aligned dipole field. The variation of the viscosity and Ohmic diffusivity, and of their ratio, the magnetic Prandtl number, has also only been explored in a limited way to date. In \citet{astoul2025interplay} we performed exploratory nonlinear MHD simulations of tidal waves in spherical shells, but many aspects of the linear response of magnetized rotating fluid bodies remain unexplored to date.

In this work, we perform a combination of linear direct numerical magnetohydrodynamic (MHD) simulations and linear boundary value calculations to analyse the tidal response in an incompressible spherical shell model of a stellar or planetary convective envelope. We build upon prior work by exploring a wide range of magnetic field strengths and configurations, including misaligned dipole fields, purely azimuthal/toroidal fields, and mixed poloidal-toroidal fields for the first time in this context. We also set out to explore a wider range of parameters than in prior work, including various spherical shell thicknesses, Ekman numbers (the ratio of viscous to Coriolis accelerations) and magnetic Prandtl numbers (ratio of kinematic viscosity to Ohmic diffusivity). We specifically focus on exploring values of the Ekman number and magnetic Prandtl number that more closely resemble expectations from turbulent values motivated by mixing-length theory \citep[e.g.][]{OL2007,Duguid2020,kapyla2020turbulent,Bekki2022,Nils2023}. 

The structure of this paper is as follows. In \S~\ref{Model}, we describe our model and numerical methods, and specifically describe the various background magnetic field configurations we consider in \S~\ref{field}. Our results are presented in \S~\ref{results}, along with our conclusions in \S~\ref{conclusions}.

\section{Model} \label{Model}

We consider a uniformly rotating spherical shell consisting of incompressible fluid, which represents a simplified model of the convective envelope of a star or planet. To determine how the magnetic field affects wave-like tidal perturbations in the fluid envelope, we assume that the entire body is permeated by a magnetic field and the fluid in the envelope is both viscous and electrically conducting. This section introduces the governing MHD equations and the numerical methods we use to solve them.

  \subsection{Linearised MHD equations for tidal flows with an imposed background magnetic field}\label{equations}
  
  A perfectly rigid inner core\footnote{Extending the model to consider a stably-stratified inner core (radiative zone) is straightforward, but this is deferred to future work. Nevertheless, our model is approximately valid for modelling the envelopes of solar-type stars as long as the core is sufficiently strongly stably stratified that it can act as a rigid boundary for the waves themselves.} of radius $r=\alpha R$ is enclosed by a homogeneous, incompressible and electrically conducting fluid shell with a free surface at $r=R$.  We assume that the envelope consists of a viscous fluid with a constant density $\rho$ (set to 1, without loss of generality), kinematic viscosity $\nu$ and Ohmic diffusivity $\eta$ (which might represent turbulent viscosity and diffusivity due to convective turbulence acting on tidal flows). The shell rotates as a solid body with a frequency $\Omega$ along the vertical unit vector $\vec{e}_z$, and is permeated by a steady background magnetic field $\vec{B}_0$. This background field $\vec{B}_0$ is assumed to be perfectly maintained by a turbulent convective dynamo, or perhaps by other processes which we do not explicitly model. This approach is justified if we are primarily interested in the effects of the field on the linear tidal response. The linearised momentum, induction, continuity equations, and solenoidal constraint on the magnetic field for the tidally excited magneto-inertial waves are given by:
  \begin{align}
   & \begin{aligned}
      \frac{\partial \vec{u}}{\partial t}+2\vec{e}_z\times \vec{u}  =& -\vec{\nabla}p+\mathrm{Le}^2(\vec{\nabla}\times \vec{B})\times \vec{B}_0+\mathrm{Le}^2(\vec{\nabla}\times \vec{B}_0)\times \vec{B} \nonumber \\
    & +\mathrm{Ek}\vec{\nabla}^2\vec{u}+\vec{f}_\mathrm{t}, 
	\label{eq:mom} 
    \end{aligned} 
    \numberthis \\
   & \frac{\partial \vec{B}}{\partial t}=\vec{\nabla}\times \left(\vec{u}\times\vec{B}_0\right)+\mathrm{Em}\vec{\nabla}^2\vec{B},
	\label{eq:intro} \\
   & \vec{\nabla}\cdot \vec{u}=0,
	\label{eq:con}  \\
   & \vec{\nabla}\cdot \vec{B}=0.
	\label{eq:Bcon}
  \end{align}
  Here, $\vec{u}$, $\vec{B}$ and $p$ are the dimensionless perturbed velocity, magnetic field, and pressure. $R$,\,$\Omega^{-1}$, and $B_{\textrm{amp}}$ (a measure of the strength of the magnetic field) are adopted as units of length, time and magnetic field, respectively. This system contains three dimensionless numbers: the Lehnert number $\mathrm{Le}$, the Ekman number $\mathrm{Ek}$ and the magnetic Ekman number $\mathrm{Em}$, which are defined as:
  \begin{equation}
    \mathrm{Le}=\frac{B_{\mathrm{amp}}}{\sqrt{\rho \mu_0}\Omega R},\quad  \mathrm{Ek}=\frac{\nu}{\Omega R^2}, \quad \mathrm{Em}=\frac{\eta }{\Omega R^2}\,.
	\label{eq:LeEkEm}
  \end{equation}
  The Lehnert number $\mathrm{Le}$ measures the strength of the background magnetic field with respect to rotation, i.e. the ratio between the Alfv\'en velocity $B_{\mathrm{amp}}/\sqrt{\rho \mu_0}$ and the rotational velocity $\Omega R$. The Ekman number $\mathrm{Ek}$ is the ratio between the rotation time-scale $\Omega^{-1}$ and the viscous time-scale $R^2/\nu$. The magnetic Ekman number $\mathrm{Em}$ is the ratio between the rotation timescale $\Omega^{-1}$ and the magnetic diffusion time-scale $R^2/\eta$. Here, $\nu$ and $\eta$ are the (constant) kinematic viscosity and Ohmic diffusivity of the fluid, respectively. $\mathrm{Ek}$ and $\mathrm{Em}$ are related by the magnetic Prandtl number 
  \begin{equation}
    \mathrm{Pm}=\frac{\mathrm{Ek}}{\mathrm{Em}}=\frac{\nu}{\eta}.
	\label{eq:Pm}
  \end{equation}
  Here, Ek may be considered to represent a turbulent viscosity motivated by values obtained from mixing-length theory \citep[usually $\mathrm{Ek}\approx10^{-5}$, e.g.,][relevant for solar-like envelopes]{OL2007,Duguid2020,astoul2022effects,Bekki2022}, and we may also expect $\eta$ to represent a turbulent value, in which case Pm may be $O(1)$ \citep[e.g.][]{kapyla2020turbulent}. The microscopic kinematic viscosities and Ohmic diffusivities in the convection zones of stars and planets are typically much smaller, with values of $\mathrm{Pm}\approx 7\times 10^{-2}$ in the lowest and densest parts of the solar convective envelope \citep[e.g.][]{Gough2007}, decreasing to $10^{-6}-10^{-5}$ for the lowest density regions nearer to the surface \citep[e.g. as we have computed from evaluating them from solar models such as those in][]{JCD1996}, with our analysis of solar models indicating a density-weighted average throughout the convective envelope of $\mathrm{Pm}\approx 8\times 10^{-3}$. The convective regions of low-mass stars generally have spatially-averaged Pm ranging from $10^{-8}$ to $10^{-2}$ \citep[e.g.~Fig.~2 in][though this does not appear to use density-weighted values]{Aug2019}, whereas the convective cores of intermediate- and high-mass stars can have Pm ranging from $10^{-1}$ to larger than unity values such as $10^2$. On the other hand, in 
  giant planets we typically expect even smaller values like in Jupiter where $\mathrm{Pm}\sim 10^{-6}$ deep in the interior and even lower values close to the surface \citep[e.g.][]{Jones2011,French2012,GW2021}. While the turbulent parameter values quoted here are achievable with our numerical computations, the microscopic values remain inaccessible. We could alternatively interpret our adopted values of $\mathrm{Ek}$ and $\mathrm{Em}$ as microscopic ones for a body that is much more viscous and much less electrically conducting than a real star or giant planet, in which case we would need to explore how the physics in our model varies with the diffusivities to extrapolate our results to astrophysical parameter regimes. Due to the uncertainties in the values of Ek and Em that we should use, even if they represent turbulent values, and to the impossibility to directly probe astrophysical parameter regimes of the microscopic diffusivities with these kinds of calculations, one of our goals is therefore to explore how the tidal response depends on Ek and Em.

  We decompose the low-frequency (relative to the dynamical frequency characteristic of surface gravity modes) tidal flow into an equilibrium/non-wavelike tide and a dynamical/wavelike tide as in \citet{ogilvie2013tides,lin2018tidal,astoul2022effects,astoul2023tidally,astoul2025interplay}. The equilibrium tide is determined by the quasi-hydrostatic adjustment of a body and its associated flow, due to the tidal and self-gravitational potentials of the perturber and the perturbed body, respectively \citep[][]{ogilvie2014tidal}. The equilibrium tidal flow satisfies a free boundary condition on the tidally-perturbed surface, whereas the wavelike tide satisfies impenetrable boundary conditions on a spherical shell, the latter of which can be demonstrated to be asymptotically consistent for low-frequency tidal forcing in linear theory. The wavelike tide is forced by an effective forcing due to the action of the Coriolis acceleration on the equilibrium tidal flow \citep[see also][]{lin2018tidal,Astoul2019}:
  \begin{equation}
    \vec{f}_\mathrm{t}=\mathrm{Re}\left\{-\vec{e}_z\times\vec{\nabla}\left [ (r^2+\frac{2}{3}\alpha^5r^{-3})Y_{2}^{2}(\theta, \phi) \right ] \frac{\mathrm{i}\omega C_t}{1-\alpha^5} \mathrm{e}^{-\mathrm{i}\omega t}\right\}.
	\label{eq:tidalforce}
  \end{equation}
  We have written Eq.~(\ref{eq:tidalforce}) using dimensionless units using spherical polar coordinates $(r,\theta,\phi)$ centred on the body, with the rotation axis along $\theta=0$. Here, $t$ is time, $Y_{2}^{2}=\frac{1}{4}\sqrt{\frac{15}{2\pi}}\sin^2\theta\,\mathrm{e}^{2\mathrm{i}\phi}$ is a quadrupolar spherical harmonic with harmonic degree $l$ and azimuthal wavenumber $m$ satisfying $l=m=2$, $\omega$ is the tidal forcing frequency, and $C_t$ is the dimensionless tidal amplitude \citep[e.g.][]{astoul2022effects}.
  
  Note that for a full sphere, when $\alpha=0$, $\vec{f}_\mathrm{t}\propto \nabla (x y \cos \omega t+(x^2-y^2)\sin \omega t)$, where $x$ and $y$ are Cartesian coordinates, and hence the effective forcing is irrotational ($\nabla \times \vec{f}_\mathrm{t}=\vec{0}$). In this case, no inertial waves are excited \citep[e.g.][]{ogilvie2013tides}, and the effective tidal forcing can be balanced by a time-dependent pressure gradient with no wavelike flow or magnetic field perturbation. The equilibrium tide is thus the exact solution. Hence, we expect no tidal dissipation of wavelike tides when $\alpha=0$ in our incompressible model (with or without a magnetic field) -- though note that this does not apply in more realistic compressible models \citep[e.g.][Vanon et al., submitted]{ogilvie2013tides}.
  
  In our study, $C_t$ is an input parameter, though for the astrophysical problem it is related to the dimensionless tidal amplitude $\epsilon=(M_2/M_1)(R/a)^3$ by\footnote{Note that we have corrected a numerical pre-factor compared to the definition of $C_t$ in AB22 and AB23.} $C_t=\sqrt{6\pi/5}(1+k_2)\epsilon$. Here, $M_2$ and $M_1$ are the masses of the perturber and perturbed body, respectively. $a$ is the orbital semimajor axis, and $k_2$ is the real part of the quadrupolar Love number (typically approximated by its hydrostatic value). In linear theory, we can set $C_t=1$ without loss of generality. Any nonzero value of $C_t$ behaves identically to any other, except differing by an overall scaling factor which is known in advance (since the tidal flow $|\boldsymbol{u}|=O(C_t)$, and hence $|\boldsymbol{B}|=O(C_t)$, the energetic quantities and dissipation rates scale as $O(C_t^2)$, so our results can be simply rescaled). The assumption of linear tides is formally appropriate if $C_t\ll 1$. However, it should be noted that nonlinear effects -- particularly those associated with short-wavelength tidal waves -- may become important even for very small $C_t$ \citep[e.g.][]{FBBO2014,astoul2022effects,astoul2023tidally,astoul2025interplay}. We defer a study of nonlinear effects to future work \citep[see also][]{astoul2025interplay}.

 To minimise the viscous and electromagnetic couplings between the fluid layer and the rigid core, and with the external medium, we use the stress-free boundary condition for the velocity $\vec{u}$ and the insulating boundary condition for the magnetic field $\vec{B}$, at both the inner and outer boundaries. The stress-free boundary condition means that the tangential component of the viscous stress should vanish:
  \begin{equation}
     u_r=\frac{\partial }{\partial r}\left(\frac{u_\theta}{r}\right)=\frac{\partial }{\partial r}\left(\frac{u_\phi}{r}\right)=0.
	\label{eq:ubc}
  \end{equation}
  The insulating boundary condition implies that no electric current flows through the boundaries:
  \begin{equation}
     \left( \vec{\nabla}\times \vec{B} \right)_r=0.
	\label{eq:Bbc}
  \end{equation}
  This is consistent with the external medium being a perfect  vacuum, but is an approximation when modelling stars with radiative cores, for example.

  \subsection{Background magnetic field configurations}\label{field}
  Since the magnetic field strengths and configurations in the interiors of stars and planets are poorly constrained, we consider various configurations and strengths of the background magnetic field to see how they affect the wavelike tidal flow. We describe these below.
  
  \subsubsection{Tilted dipole field}
  We first consider a dipolar magnetic field that is tilted away from the rotation axis $\theta=0$ by an angle $\theta_0$ in the meridional plane. This is in general a tilted/misaligned dipolar field -- though it is an aligned dipole if $\theta_0=0$. It is defined as
  \begin{align}
   &B_{0,r}=\frac{1}{r^3}\left(\mathrm{cos}\theta_0 \,\mathrm{cos}\theta+\mathrm{sin}\theta_0\,\mathrm{sin}\theta \,\mathrm{cos}\phi\right) ,\nonumber \\
   &B_{0,\theta}=\frac{1}{2\,r^3}\left(\mathrm{cos}\theta_0\,\mathrm{sin}\theta -\mathrm{sin}\theta_0 \,\mathrm{cos}\theta \,\mathrm{cos} \phi\right),\nonumber \\
   &B_{0,\phi}=\frac{1}{2\,r^3} \mathrm{sin}\theta_0\,\mathrm{sin}\phi.
    \label{eq:dipolar}  
  \end{align}
  where we recall also that its magnitude has already been incorporated into the Lehnert number, and similarly $r$ here is already nondimensionalised by the outer radius of the shell.
  This equation recovers the expression for an aligned dipolar field when $\theta_0=0$, which has been studied by \citet{RR2003} and \citet{lin2018tidal}, and agrees with the horizontal dependence of the tilted dipole considered by \citet{WeiGoodman2015}, although the latter considered a free decay mode for the radial profile, whereas this expression is for a ``core-generated'' dipolar field decaying as $r^{-3}$. Note that the misaligned dipole (when $\theta_0\ne n \pi$ for any integer $n$) is not axisymmetric, unlike the aligned dipole, and that it possesses an azimuthal structure with wavenumber $m=1$. Hence, linear tidal forcing with $m=2$ will excite multiple $m$ components of the field and flow, in general. 
  
  The motivation for considering a tilted dipolar field is that the magnetic axes of some planets are observed to be misaligned with their rotation axes, e.g., the tilt of the dipole component relative to the rotation axis is approximately $11^{\circ}$ for the Earth and about $9.5^{\circ}$ for Jupiter, while Saturn’s field is nearly aligned with its spin axis. Uranus and Neptune show much larger tilts \citep[$\sim 59^{\circ}$ and $\sim 47^{\circ}$, respectively, e.g.][]{schubert2011planetary}. Moreover, the host stars of close-in exoplanets can also feature a tilted dipolar magnetic field, such as  TAP 26 and Kepler-78A, which are young K-type stars whose magnetic axes are tilted by an angle of $40^{\circ}$ and $50^{\circ}$ with respect to the rotation axis, respectively \citep[][]{MD2016,YD2017}. 
    
  We consider an aligned dipolar field ($\theta_0=0$) and two tilted dipolar fields ($\theta_0=\pi/4,\,\theta_0=\pi/2 $), whose three-dimensional magnetic field lines are shown in Fig.~\ref{fig:B0}(\textit{c,\,d,\,e}). We do not need to consider larger tilt angles with $\theta_0\geq \pi/2$. This is because the tidal flow driven by the tidal force as in Eq.~\eqref{eq:tidalforce} is mirror-symmetric with respect to the equatorial plane, such that an angle $\theta_0$ has the same effect on the dynamics as an angle $\pi-\theta_0$. In addition, we have the usual symmetry in MHD that the dynamics is invariant under the transformation $\boldsymbol{B}\to -\boldsymbol{B}$. Similarly, we do not need to consider any tilt angle for the field in the $\phi$ direction (out of the meridional plane).
  
\subsubsection{Azimuthal ``Malkus field"}
  As our first alternative to a dipole, we also consider a Malkus field \citep{Malkus1967}, as presented in Fig.~\ref{fig:B0}(\textit{a}). This is an axisymmetric, purely toroidal magnetic field that is defined as  
  \begin{align}
   \vec{B}_{0}=S_m \,r \,\mathrm{sin}\theta \boldsymbol{e}_\phi.
    \label{eq:malkus}  
  \end{align}
  This has a uniform current $\nabla \times \vec{B}_0=S_m \vec{e}_z$, everywhere in space. While this field is particularly idealised, it allows us to explore the effects of a toroidal field on the tidal response. 

  As a way to more fairly compare the effects of this field with the dipolar field above, we have introduced a scaling factor $S_m$ to ensure that the volume integral of the dimensionless magnetic energy (for a given $\mathrm{Le}$) in the shell is identical between the dipolar-type field and the Malkus field. Mathematically, this means that $\left<\left(r^{-3}\mathrm{cos}\theta\, \hat{\vec{r}}+\frac{1}{2}r^{-3}\mathrm{sin}\theta\, \hat{\vec{\theta}}\right)^2 \right>=\left< (S_m \,r\,\mathrm{sin}\theta)^2 \right>$, where $\left< \cdot \right>$ indicates a volume integral over the shell. Here, $S_m=3.005$ for $\alpha=0.5$.

  \subsubsection{Mixed poloidal-toroidal ``Prendergast field''}
  Our third type of magnetic field considered is an axisymmetric, mixed poloidal and toroidal field configuration referred to as a 
  ``Prendergast magnetic field" \citep[e.g.][]{prendergast1956equilibrium,Loi2019,kaufman2022stability}, as presented in Fig.~\ref{fig:B0}(\textit{b}). This is a magnetic field in magneto-hydrostatic equilibrium that is often considered as a suitable form of a ``fossil field'' in a stellar radiative zone, which may persist over stellar evolutionary timescales \citep[][for its compressible generalisation]{braithwaite2006stable,DM2010a,DM2010b,DB2010}. We consider a form of this field with closed magnetic field lines confined within the outer boundary of the star or planet.
  
  This field is defined as 
  \begin{align}
   &B_{0,r}=\frac{2}{r^2}\Psi (r)\,\mathrm{cos}\theta, 
   \; 
   B_{0,\theta}=-\frac{1}{r}{\Psi}'(r)\,\mathrm{sin}\theta,\; 
   B_{0,\phi}=-\frac{\lambda}{r}{\Psi}(r)\,\mathrm{sin}\theta.
    \label{eq:BPrend}  
  \end{align} 
  This is a non-singular solution to the balance,
  \begin{align}
     \lambda \vec{B}_0+\vec{\nabla}\times \vec{B}_0=-S_p\,\beta\, r \,\mathrm{sin}(\theta)\hat{\vec{\phi}}
    \label{eq:BPrend1}  
  \end{align} 
  with $\vec{B}_0=\vec{0}$ at the outer boundary to ensure that the magnetic field is confined within the body. This is a magneto-hydrostatic equilibrium condition for the magnetic field that is axisymmetrically forced by the baroclinic torque (arising from buoyancy forces and differential rotation) on the right-hand side. $\beta$ determines the field amplitude (inherited from the mechanical driving forces) and $S_p=10.082$ is a scaling factor that is designed to act like $S_m$ above. $\Psi(r)$ is defined as 
  \begin{align}
     \Psi(r)=\frac{S_p\beta}{\lambda^2}\left(r^2-r\frac{j_1(\lambda r)}{j_1(\lambda)}\right),
    \label{eq:bassel}  
  \end{align} 
  where $j_1(\xi )=\mathrm{sin}(\xi)/\xi^2-\mathrm{cos}(\xi)/\xi$ is a spherical Bessel function of the first kind. The entire field vanishes at the outer boundary provided ${\Psi}(r=1)={\Psi}'(r=1)=0$, which leads to the eigenvalue condition $\tan(\lambda)=3\lambda/(3-\lambda^2)$. We choose the lowest energy solution with $\lambda \approx 5.76346$ \citep[see also][]{kaufman2022stability}. 
  
  Note that \citet{kaufman2022stability} found this field to be unstable to a resistive instability on very long timescales, which may complicate our analysis of its effects on tidal waves. However, for our purposes, this instability is very slowly growing relative to the dynamical timescales associated with tidal wave propagation and dissipation, and therefore does not significantly affect the solutions considered here. We have performed some linear simulations with initial random noise (without tides) to analyse the stability of this field configuration and have observed it to be stable for the parameters we have considered, with no instability evident within $10^4\Omega^{-1}$.

  \subsubsection{Axisymmetric free-decay poloidal dipole and quadrupole fields}
\begin{figure*}
	\includegraphics[width=2\columnwidth]{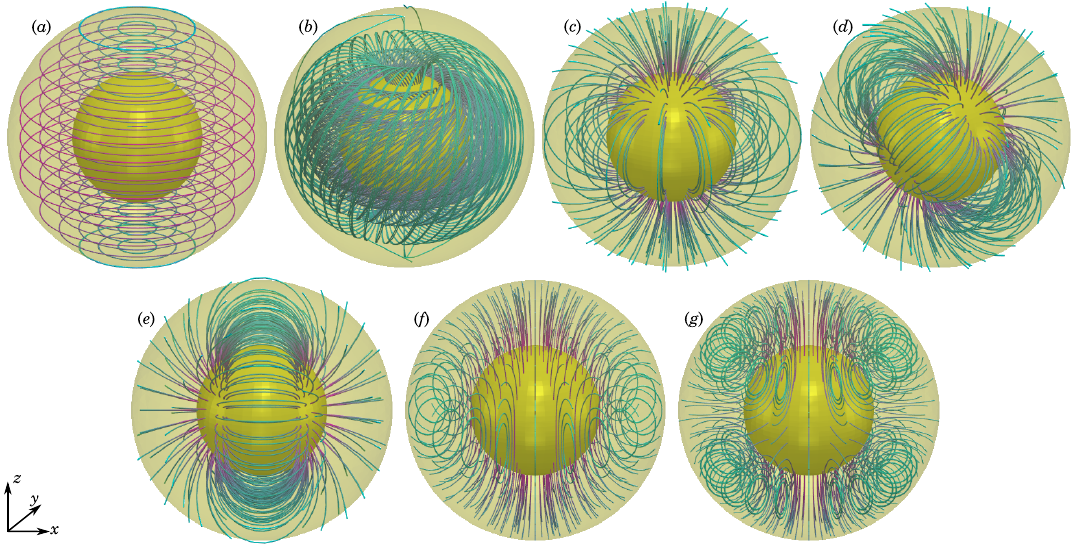}
    \caption{Illustration of the three-dimensional magnetic field lines for each of the background magnetic fields we consider. $(a)$ Malkus field; $(b)$ Prendergast field; $(c)$ Aligned dipolar field; $(d)$ tilted dipolar field with $\theta_0=\pi/4$; $(e)$ tilted dipolar field with $\theta_0=\pi/2$. $(f)$ Axisymmetric free-decay poloidal dipole field. $(g)$ Axisymmetric free-decay quadrupole field. Colours denote the strength of the magnetic field, representing the strongest values as pink and the weakest as blue.}
    \label{fig:B0}
  \end{figure*}

  Finally, we also consider two examples of axisymmetric poloidal free-decay fields, which have spatial structures that are eigenmodes of the magnetic diffusion operator. We consider both dipolar $(l=1)$ and quadrupolar $(l=2)$  free-decay modes to assess the sensitivity of the tidal magnetic coupling to the large-scale magnetic geometry, which are labelled as $l=1$ and $l=2$ for simplicity in our results below. The three-dimensional magnetic field lines of both of these are shown in Fig.~\ref{fig:B0}(\textit{f,g}). Unlike the dipolar field considered above, which represents a radially decaying field outside the core, these fields are eigenfunctions of the magnetic diffusion equation in the spherical shell with our adopted insulating boundary conditions. They introduce no spurious lengthscales, and they provide a clean and computationally efficient reference state for analysing the effects of the magnetic field on the tidal response. These versions of dipolar and quadrupolar fields are less spatially localised towards the inner boundary than the previously defined dipole considered above.

  To derive equations describing these two fields, we start with the poloidal-toroidal decomposition for the magnetic field:
  \begin{align}
   \vec{B}=\vec{\nabla}\times\vec{\nabla}\times (g \vec{\hat{r}})+\vec{\nabla}\times (h \vec{\hat{r}}),
    \label{eq:ptdecomposation}  
  \end{align}
  where $g$ and $h$ are the poloidal and toroidal scalar fields.
  As we only consider the axisymmetric poloidal component, we have $h=0$, azimuthal wave number $m=0 $, and $g(r,\theta,\phi)=g_{1}(r)Y_1^0(\theta,\phi)$ for the dipolar field and $g(r,\theta,\phi)=g_{2}(r)Y_2^0(\theta,\phi)$ for the quadrupolar field. Here, $Y_1^0(\theta,\phi)=\sqrt{\frac{3}{4\pi}}\mathrm{cos}\theta$ and $Y_2^0(\theta,\phi)=\sqrt{\frac{5}{16\pi}}(3\mathrm{cos}^2\theta-1)$ are orthonormalised spherical harmonics. 

  The magnetic diffusion equation is
  \begin{align}
   \frac{\partial \vec{B}}{\partial t}=\mathrm{Em}\vec{\nabla}^2 \vec{B},
    \label{eq:mde}  
  \end{align}
  which can be projected onto individual harmonic modes to obtain the governing equation for each $g_l$. 
  For eigenmodes that decay exponentially as 
     $g_l(r,t)=\hat{g}_l(r)e^{-\kappa t}$,
  with $\kappa>0$,   we obtain the equation governing $\hat{g}_l(r)$ \citep[][]{AS1972,M1978}:
  \begin{align}
     \left ( \frac{\mathrm{d}^2\hat{g}_l}{\mathrm{d}r^2}-\frac{l(l+1)}{r^2}\hat{g}_l\right ) +k^2 \hat{g}_l=0, \,\,\, \text{where} \,\,\, k^2=\frac{\kappa}{\mathrm{Em}}.
    \label{eq:glODE}  
  \end{align}
  The solution for each $l$ involves a linear combination of spherical Bessel functions of the first and second kinds, of integral order,
  respectively, $j_l(kr)$ and $y_l(kr)$:
  \begin{align}
     \hat{g}_l(r)=
     A j_{l}(kr)+B y_{l}(kr).
    \label{eq:glsolution}  
  \end{align}
   Here, $A$ and $B$ are complex constants that are determined by the insulating boundary conditions at the inner and outer spherical surfaces:
   \begin{align}
     \frac{\partial \hat{g}_l}{\partial r}-\frac{l+1}{r} \hat{g}_l=0,\, \,\,\, \text{at} \,\,\,r=\alpha,
    \label{eq:bcinner}  
   \end{align}
   and
  \begin{align}
     \frac{\partial \hat{g}_l}{\partial r}+\frac{l}{r} \hat{g}_l=0,\, \,\,\, \text{at} \,\,\,r=1.
    \label{eq:bcout}  
  \end{align}
   When combining equations \eqref{eq:glsolution}--\eqref{eq:bcout}, we can obtain the eigenvalue $k$ and eigenfunctions $\hat{g}_l(r)$ for a given spherical harmonic degree $l$ at fixed Em. Here, we pick the slowest-decay eigenvalue (smallest value of $\kappa$) and its corresponding eigenfunction. 
  
  For $l=1,\alpha=0.5$, the slowest-decay eigenvalue $k_1=2.9939$ and $A/B\approx4.6484$ (using $A=4.6484$ and $B=1$ without loss of generality as we rescale the field amplitude further below). The axisymmetric poloidal dipole field is given by 
  \begin{align}
     & B_{0,r}=\frac{2}{r^2}\hat{g}_{1}(r)\sqrt{\frac{3}{4\pi}}\mathrm{cos}\theta,\nonumber \\
     & B_{0,\theta}=-\frac{1}{r}\hat{g}'_{1}(r)\sqrt{\frac{3}{4\pi}}\mathrm{sin}\theta,\nonumber\\
     & B_{0,\phi}=0.
    \label{eq:freedipo}  
  \end{align}
  Here, 
    \begin{align}
     & \hat{g}_{1}(r)=
     S_d\left[A j_{1}(k_1r)+B y_{1}(k_1r)\right],\\
     & 
     j_1(x)=\frac{\mathrm{sin}x}{x^2}-\frac{\mathrm{cos}x}{x},\;\; y_{1}(x)=-\frac{\mathrm{cos}x}{x^2}-\frac{\mathrm{sin}x}{x}.
    \label{eq:freedipo1}  
  \end{align}
  This has non-vanishing current in the shell, $\vec{\nabla} \times \boldsymbol{B}_0=\sqrt{3/(4\pi)}(k_1^2/r)\hat{g}_{1}(r)\sin\theta\boldsymbol{e}_\phi$.
  
  For $l=2,\alpha=0.5$, the slowest-decay eigenvalue $k_2=4.3733$ and $A/B\approx7.9951$ (We use $A=7.9951$ and $B=1$ without loss of generality). The axisymmetric poloidal quadrupole field is given by 
  \begin{align}
     & B_{0,r}=\frac{6}{r^2}\hat{g}_{2}(r)\sqrt{\frac{5}{16\pi}}\left ( 3\mathrm{cos}^2\theta-1 \right ),\nonumber \\
     & B_{0,\theta}=-\frac{6}{r}\hat{g}'_{2}(r)\sqrt{\frac{5}{16\pi}} \mathrm{cos}\theta\,\mathrm{sin}\theta , \nonumber \\
     & B_{0,\phi}=0.
    \label{eq:freeqdipo}  
  \end{align}
  Here, 
    \begin{align}
     & \hat{g}_{2}(r)=
     S_{qd}\left(A j_2(k_2r)+B y_{2}(k_2r)\right),\\
     & j_{2}(x)= \left(\frac{3}{x^2}-1 \right)\frac{\mathrm{sin}x}{x}-\frac{3}{x^2}\mathrm{cos}x,\\
     & y_{2}(x)= -\left (\frac{3}{x^2}-1 \right )\frac{\mathrm{cos}x}{x} -\frac{3}{x^2}\mathrm{sin}x .
    \label{eq:freeqdipo1}  
  \end{align}
  This has non-vanishing current in the shell, $\vec{\nabla} \times \boldsymbol{B}_0=\sqrt{5/(16\pi)}(6k_2^2/r)\hat{g}_{2}(r)\cos\theta\sin\theta\boldsymbol{e}_\phi$.
  
  We define the factors $S_d=0.798523$ and $S_{qd}=0.2301$
  to equalize the volume-integrated magnetic energies of our various configurations for a given Le, in a similar way to the parameter $S_m$ above.
  
  \subsection{Energy balance}
  The energy balance satisfied by the tidal flow and magnetic field perturbations can be obtained by deriving an evolutionary equation for the total energy. To obtain this, we first take dot products of equation \eqref{eq:mom} with $\boldsymbol{u}$ and equation \eqref{eq:intro} with $\mathrm{Le}^2\boldsymbol{B}$. Then, summing the two after performing volume integration, we obtain the governing equation for the total (kinetic + magnetic) energy
  \begin{align}
      & \frac{\partial }{\partial t}\left (  \left< \frac{1}{2}|\vec{u}|^2\right>+\left<\frac{1}{2}\mathrm{Le}^2|\vec{B}|^2\right>\right )=P_t-D_\mathrm{vis}-D_\mathrm{ohm}-F_p+W_\mathrm{lf}.
    \label{eq:energy}  
  \end{align}
  Here, $P_\mathrm{t}=\left<\vec{u}\cdot \vec{f}_\mathrm{t}\right>$ is the work done by the tidal force, which quantifies the tidal energy transfers, 
  \begin{align}
  D_{\mathrm{vis}}&=2\,\mathrm{Ek} \left< e_{ij}e_{ij} \right>, \\
  D_{\mathrm{ohm}}&=\mathrm{Le^2Em}\left<(\vec{\nabla} \times \vec{B})^2\right>,
  \end{align}
  are the viscous and Ohmic dissipation rates, $e_{ij}=(1/2)(\partial_i u_j+\partial_j u_i)$ is the rate-of-strain tensor, and 
  \begin{align}
   F_p&=-\mathrm{Le^2}\left< \vec{\nabla}\cdot \left(\left(\vec{u}\times\vec{B}_0\right)\times \vec{B} \right )\right>,\\
   W_{\mathrm{lf}}&=\mathrm{Le^2}\left< \vec{u}\cdot \left( \left(\vec{\nabla}\times\vec{B}_0\right)\times \vec{B} \right)\right>,
   \end{align}
   represent the sum of Poynting and diffusive fluxes, quantifying the transfer of magnetic energy through the inner and outer boundaries, and the work done by part of the Lorentz force, respectively. 
   Periodic solutions to the boundary value problem that vary in time as $\mathrm{e}^{-\mathrm{i}\omega t}$ satisfy energy balance such that the right hand side of Eq.~(\ref{eq:energy}) vanishes, so that
  \begin{align}
     -D_{\mathrm{vis}}-D_{\mathrm{ohm}}-F_p+W_{\mathrm{lf}}+P_\mathrm{t}=0.
    \label{eq:energy1}  
  \end{align}
  In our simulations that solve the initial value problem, this steady state response is also attained after $O(10^4\Omega^{-1})$ for the parameters we have considered. This provides an excellent check of the consistency of our numerical results. We have verified that it has been satisfied to within a fraction of $1\%$ in each of the numerical calculations in this paper.
   
  For the radially decaying dipolar-type magnetic field, $W_{\mathrm{lf}}=0$ because $\vec{\nabla}\times \vec{B}_0=\vec{0}$. 
  But it is not the case for other fields, which are not current-free, and hence $W_{\mathrm{lf}}\ne 0$. The Poynting flux is non-zero even for an aligned dipole field, but it has always been found to be negligibly small for all of the background magnetic fields that we consider. For the calculations with the aligned dipole and tilted dipole fields, we have found that the dominant balance is $P_\mathrm{t}\approx D_\mathrm{vis}+D_\mathrm{ohm}$. On the other hand, $W_\mathrm{lf}$ is non-negligible for the other non-current free fields, so the dominant balance in those is $P_\mathrm{t}\approx D_{\mathrm{vis}}+D_{\mathrm{ohm}}-W_\mathrm{lf}$. The maximum fractional contribution of $W_{\mathrm{lf}}$ relative to $P_{\mathrm{t}}$ can be as high as 40\% for the largest Le that we consider, while that of $F_p$ is typically below 1\%. 
  
  \subsection{Numerical methods}
  We solve the system of equations (\ref{eq:mom}--\ref{eq:Bcon}), with boundary conditions (\ref{eq:ubc}) and (\ref{eq:Bbc}), as a two-dimensional boundary value problem for the azimuthal wavenumber $m=2$ tidal response with Dedalus3 \citep[][]{burns2020dedalus} for the axisymmetric magnetic fields (aligned dipolar, Malkus, Prendergast and free-decay poloidal dipole and quadrupole fields). For the tilted dipolar fields ($\theta_0=\pi/4, \pi/2$), we instead solve the system as an initial value problem with the 3D pseudo-spectral MHD code MagIC (version 6.2). This is because these fields are non-axisymmetric (with $m=1$ structure), so constructing the matrices for the corresponding boundary value problem (which is inherently three-dimensional) would be more challenging as the different azimuthal wavenumbers $m$ would be coupled by the interaction with the magnetic field. Solving these cases as an initial value problem instead avoids the requirement of constructing very large matrices when solving the three-dimensional boundary value problem, which would require prohibitive memory to store. However, using MagIC is more computationally intensive (in terms of run time) for a given problem to reach the desired solution, because we must time-step until a steady state is reached. We have successfully performed several cross-checks of the two codes for axisymmetric problems to ensure that they accurately agree with each another, therefore verifying our implementation.
  
  Both methods employ Chebyshev polynomials in the radial direction (a $\tau$-method using Chebyshev polynomials of both the first and second kinds in radius in Dedalus and Chebyshev collocation on the Chebyshev-Gauss-Lobatto nodes in MagIC) and a spherical harmonic decomposition in the azimuthal and latitudinal directions. We have implemented scripts in Dedalus that solve the governing equations directly in terms of the velocity and magnetic fields (and pressure), whereas MagIC employs a poloidal–toroidal decomposition for both the flow and field, which enforces the solenoidal constraints and reduces the number of variables. Details of the two codes can be found at \url{https://dedalus-project.readthedocs.io/en/latest/} and \url{https://magic-sph.github.io/numerics.html}, respectively.  For MagIC, our simulations have been performed until a steady state for the tidal power and dissipation has been reached (in most cases), which typically takes $O(10^4\Omega^{-1})$. We use a CNAB2 scheme (second-order scheme assembled from the combination of a Crank-Nicolson method for the implicit terms and a second-order Adams-Bashforth method for the explicit terms) for time integration, with an adaptive time-step satisfying a Courant-Friedrichs-Lewy condition with maximum $\mathrm{d}t=5\times10^{-3}$ for $\mathrm{Le=10^{-3}}$, $\mathrm{d}t=10^{-3}$ for $\mathrm{Le=10^{-2}}$ and $\mathrm{d}t=3\times 10^{-4}$ for $\mathrm{Le=5\times 10^{-2}}$. The spatial resolution used in the simulations varies from case to case, but we ensure that there is at least a three orders of magnitude drop in the kinetic and magnetic energy spectra for each component when the simulations reach a steady state. We have found that cases with stronger magnetic fields tend to require higher spatial resolution for the same parameters otherwise, which is ultimately due to the stronger coupling between spherical harmonic degrees -- while each of the axisymmetric fields couple $\ell\to \ell\pm 1$, Le controls the strength of these couplings due to the magnetic field. For the cases with $\mathrm{Le}=10^{-3}$, the typical resolution is $n_{r,\max}=65$ (number of radial grid points), $l_{\max}=64$ (maximum spherical harmonic degree), and $m_{\max}=24$ (maximum spherical harmonic order), requiring approximately 20 CPU hours per simulation. For $\mathrm{Le}=10^{-2}$, the resolution is increased to $n_{r,\max}=81$, $l_{\max}=96$, and $m_{\max}=42$, with a computational cost of about 100 CPU hours per simulation. For the higher magnetic field strength, $\mathrm{Le}=5 \times 10^{-2}$, the resolution is further refined to $n_{r,\max}=81$, $l_{\max}=128$, and $m_{\max}=60$, leading to a cost of approximately 300 CPU hours per simulation. On the other hand, the linear boundary value problems solved with Dedalus typically take a few minutes each\footnote{We used a resolution of $N_r=150$ points in radius and $N_\theta=150$ in latitude for $\mathrm{Le}=10^{-2}$, $\mathrm{Ek}=10^{-5}$, $N_r=N_\theta=180$ for $\mathrm{Le}=5\times 10^{-2}$, $\mathrm{Ek}=10^{-5}$ and $\mathrm{Le}=10^{-2}$, $\mathrm{Ek}=10^{-6}$, and $N_r=N_\theta=210$ for $\mathrm{Le}=5\times 10^{-2}$, $\mathrm{Ek}=10^{-6}$. These were found to be adequate by testing the volume-integrated quantities for convergence as the resolution was varied.}. Hence, solving the initial value problem (with MagIC) is much more computationally intensive than the boundary value solves (with Dedalus).
  
\section{Results}\label{results}

We now present our results obtained using the model and numerical methods described in \S~\ref{Model}.

 \subsection{Tidal flows at $\omega=1.1$ for various field strengths and configurations}
\begin{figure}
	   \includegraphics[width=\columnwidth]{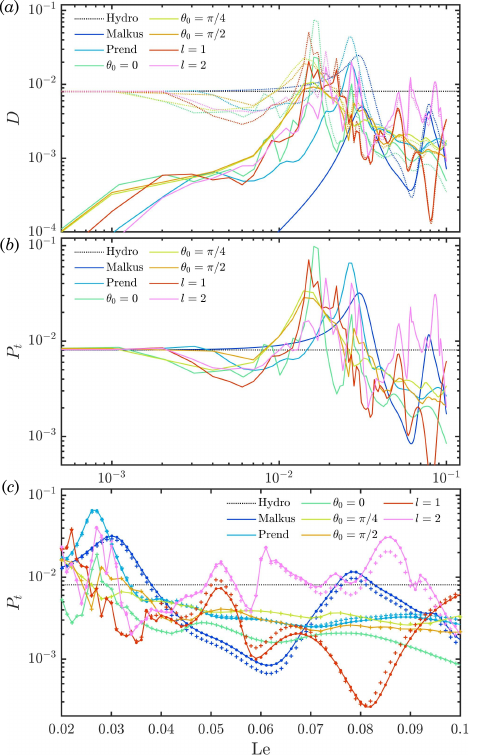}
       \caption{$(a)$ Ohmic dissipation $D_{\mathrm{ohm}}$ (solid lines), viscous dissipation $D_{\mathrm{vis}}$ (dotted lines) and $(b)$ tidal power $P_\mathrm{t}$ as a function of Le considering various background magnetic fields at $\alpha=0.5,\, \omega=1.1,\,\mathrm{Pm}=1 \,, \mathrm{Ek}=10^{-5}$. The lines with $\theta_0=0,\, \pi/4,\, \pi/2$ indicate the aligned dipolar field, and tilted dipolar fields with an angle of $\pi/4$ and $\pi/2$, respectively. $(c)$ Tidal power $P_\mathrm{t}$ and the associated energy balance according to Eq.~(\ref{eq:energy1}). Solid lines, crosses, and dots in (c) represent $P_\mathrm{t}$, $D_{\mathrm{ohm}}+D_{\mathrm{vis}}$ and $D_{\mathrm{ohm}}+D_{\mathrm{vis}}+W_\mathrm{lf}$, respectively.}
       \label{fig:Ptle}
    \end{figure}

  We begin by presenting results at a fixed tidal frequency $\omega=1.1$ and $\alpha=0.5, \mathrm{Ek}=10^{-5}, \mathrm{Pm}=1$, for different background magnetic field configurations and field strengths (values of $\mathrm{Le}$). This specific frequency was chosen because it has been widely explored in prior work studying tidally-driven inertial waves \citep[e.g.][]{lin2018tidal,astoul2022effects}. Fig.~\ref{fig:Ptle} shows the dependence of Ohmic dissipation $D_{\mathrm{ohm}}$, viscous dissipation $D_{\mathrm{vis}}$, and tidal power $P_\mathrm{t}$ on Le. As Le is increased above $10^{-3}$, the viscous dissipation gradually departs from its hydrodynamical value ($\mathrm{Le}=0$), indicating that magnetic effects have become increasingly important in modifying the tidal flow. Meanwhile, Ohmic dissipation grows from a negligible level to reach a magnitude comparable to the viscous dissipation by $\mathrm{Le}\approx 10^{-2}$. Our observation that the Ohmic and viscous dissipation rates become comparable for the largest Le contrasts with the findings of \citet{lin2018tidal}, where Ohmic dissipation was reported to dominate the total dissipation at relatively large values of Le closer to one. This discrepancy is likely attributable to the much larger Pm that we have considered (they adopted a value of $\mathrm{Pm}=10^{-4}$, which can be thought to represent a microscopic value in a giant planet, but it is much smaller than either the turbulent or microscopic values in stellar interiors), and possibly also to the larger Ekman numbers we explore here. Our values are motivated by the enhanced (turbulent) viscosity and Ohmic diffusivity associated with convective motions. These results suggest that convective turbulence can contribute significantly to viscous dissipation of tidal waves, such that it remains comparable with Ohmic dissipation even at large Le for the range of field strengths that we have considered.

  Different background magnetic field configurations exert distinct influences on tidal dissipation (quantified by $P_\mathrm{t}$), as illustrated in Fig.~\ref{fig:Ptle}(\textit{b}). Aligned and tilted dipolar fields ($\theta_0=0,\pi/4$) begin to significantly modify the tidal response at relatively small values of $\mathrm{Le}=10^{-3}$. In comparison, the free-decay poloidal dipole and quadrupole fields ($l=1,2$) affect the tidal power slightly later at $\mathrm{Le}=2\times10^{-3}$, while the Prendergast field (Prend) and strongly tilted dipolar field ($\theta_0=\pi/2$) lead to a noticeable departure from the hydrodynamical value only at $\mathrm{Le}=4\times 10^{-3}$. The Malkus field, on the other hand, only becomes influential at much larger field strengths, around $\mathrm{Le}=8\times 10^{-3}$. These trends suggest that magnetic configurations with stronger poloidal components exert a more pronounced influence on tidal flows for a given Le. In particular, the purely toroidal Malkus field produces a relatively smooth variation in tidal power with increasing $\mathrm{Le}$, whereas dipolar fields and the free-decay poloidal dipole and quadrupole fields induce more fluctuating responses as Le varies. Tilted dipolar fields also substantially influence the dissipation for a similar range of Le as the aligned dipole, but the resulting response differs markedly from the aligned dipole.

  Fig.~\ref{fig:Ptle}(\textit{c}) presents the tidal power and its calculation using the energy balance given by Eq.~(\ref{eq:energy1}) for the largest Le considered. For the Malkus field, Prendergast field, and the free-decay poloidal dipole and quadrupole fields, the tidal power $(P_\mathrm{t})$ is not accurately balanced by the total dissipation $(D_{\mathrm{ohm}}+D_{\mathrm{vis}})$. However, excellent agreement is recovered once the work done by the Lorentz force, $W_\mathrm{lf}$, is taken into account. The sign of $W_\mathrm{lf}$ differs among magnetic configurations: it is positive 
  for the Malkus field and negative
  for the Prendergast field, indicating that the background magnetic field can either supply energy to or extract energy from the tidal flow. In contrast, for dipolar-type fields, the tidal power remains largely consistent with the total dissipation, suggesting that the contribution from the Poynting flux is negligible in these cases.
  The work $W_{\mathrm{lf}}$ associated with the free-decay poloidal dipole and quadrupole fields is also significant relative to the tidal power $P_{\mathrm{t}}$. This effect is particularly pronounced for the free-decay quadrupole field, with the maximum ratio $W_{\mathrm{lf}} / P_{\mathrm{t}}$ reaching up to 34.7\%.

\begin{figure}
	   \includegraphics[width=\columnwidth]{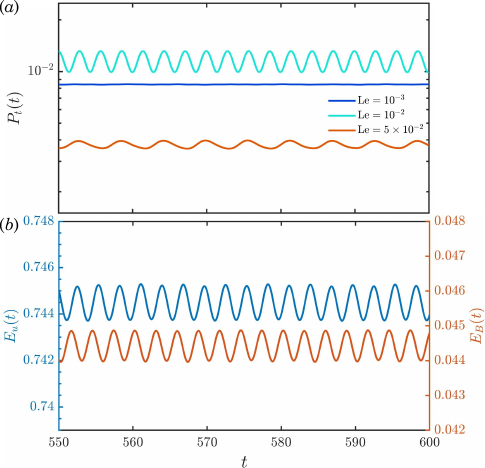}
       \caption{$(a)$ Time evolution of tidal power for the periodic steady-state tidal flow for the case of a tilted dipolar field ($\theta_0=\pi/4$) at various Le. $(b)$ Time evolution of kinetic energy ($E_u=\left< |\boldsymbol{u}|^2/2\right>$) and magnetic energy ($E_B=\left< \mathrm{Le}^2|\boldsymbol{B}|^2/2\right>$) at $\mathrm{Le}=10^{-2}$. 
       }
       \label{fig:rpo}
    \end{figure}

\begin{figure}
	   \includegraphics[width=\columnwidth]{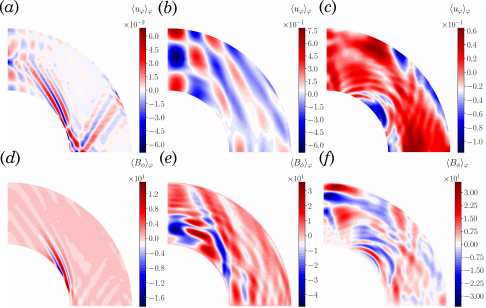}
       \caption{Azimuthally-averaged azimuthal velocity and azimuthal magnetic field in the meridional plane. $\left < \cdot \right >_{\phi}$ indicates the azimuthal average operation on the final periodic (but otherwise steady) tidal flow in the presence of a tilted dipolar field ($\theta_0=\pi/4$) at various Le and $\omega=1.1\,, \mathrm{Ek}=10^{-5}\,, \alpha=0.5\,, \mathrm{Pm}=1$. $(a,d)$ $\mathrm{Le}=10^{-3}$; $(b,e)$ $\mathrm{Le}=10^{-2}$;$(c,f)$ $\mathrm{Le}=5\times 10^{-2}$.}
       \label{fig:zonalflow}
    \end{figure}

  For all of the axisymmetric magnetic field configurations, the final resulting tidal flow is steady. In contrast, this is no longer the case for the tilted dipolar fields. As shown in Fig.~\ref{fig:rpo}(\textit{a}), which presents the time evolution of the tidal power at $\mathrm{Le}=10^{-3}\,,10^{-2}\,,5\times10^{-2}$  for a tilted dipolar field ($\theta_0=\pi/4$), the tidal power exhibits clear periodic fluctuations. Moreover, the amplitude of these fluctuations increases initially at $\mathrm{Le}=10^{-2}$, before decreasing at $\mathrm{Le}=5\times10^{-2}$, in agreement with the trend in tidal dissipation as a function of $\mathrm{Le}$ shown in Figure~\ref{fig:Ptle}(\textit{b}). This oscillatory behaviour is linked to the fluctuations of the perturbed velocity and magnetic fields. The angle between the perturbed velocity field and the background magnetic field varies periodically, which causes the induction term $\nabla \times (\vec{u} \times \vec{B}_0)$ and Lorentz force $(\nabla \times \vec{B})\times \vec{B}_0$ to oscillate in time, thereby leading to periodic variations in both the perturbed velocity and magnetic fields, and their corresponding energies. 
  This is illustrated in Fig.~\ref{fig:rpo}(\textit{b}), which shows the temporal evolution of the kinetic energy ($E_u=\left< |\boldsymbol{u}|^2/2\right>$) and magnetic energy ($E_B=\left< \mathrm{Le}^2|\boldsymbol{B}|^2/2\right>$) at $\mathrm{Le}=10^{-2}$. The two energies are fluctuating and strongly anti-correlated: an increase in kinetic energy corresponds closely to a decrease in magnetic energy, and vice versa. 
  Another notable feature is that tilted dipolar fields can generate oscillatory zonal flows even within a linear framework, as we show in Fig.~\ref{fig:zonalflow} for ($\theta_0=\pi/4$). This is impossible for axisymmetric fields and $m=2$ tidal forcing in linear theory, although zonal flows can be generated nonlinearly \citep[e.g.][]{tilgner2007,FBBO2014,astoul2022effects}. On the other hand, this is possible in linear theory for tilted dipolar fields because they have $m=1$ symmetry, and therefore different $m$ components of the flow and field are coupled despite tidal forcing being restricted to $m=2$. As Le is increased from $\mathrm{Le}=10^{-3}$ to $\gtrsim 10^{-2}$, the zonal flow becomes stronger. For $\mathrm{Le}=10^{-3}$, this zonal flow is primarily associated with the presence of $m=0$ inertial waves, as indicated by the nearly straight wave beams. For $\mathrm{Le}=10^{-2}$, the flow becomes more complex and is dominated by fast magneto-inertial waves, resulting from the interaction between inertial waves and the magnetic field. For $\mathrm{Le}=5 \times 10^{-2}$, the zonal flow is mainly governed by Alfvén-like waves, which propagate along the magnetic field lines. A similar behaviour is also observed for a tilted dipolar field with $\theta_0=\pi/2$. It arises because the tilted dipole has $m=1$ symmetry, so the linear operator couples the $m=2$ tidal response to other azimuthal harmonics, including $m=0$. The resulting oscillatory “zonal flow” is therefore an axisymmetric component of the periodic linear solution, not a nonlinearly-generated mean flow like those in \citet{astoul2022effects}.


\begin{figure*}
	   \includegraphics[width=2\columnwidth]{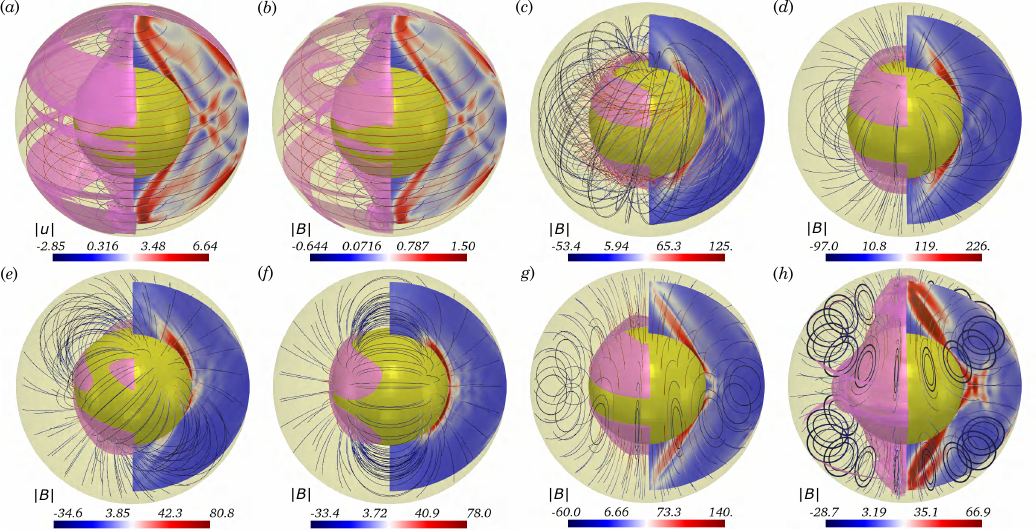}
       \caption{Pseudo-colour maps in the meridional plane and iso-surfaces in the shell for $(a)$ velocity perturbation $|\vec{u}|$ and $(b-h)$ magnetic perturbations $|\vec{B}|$ when considering different background magnetic fields $\vec{B}_0$ at $\mathrm{Le}=10^{-3}$ and $\omega=1.1\,, \mathrm{Ek}=10^{-5}\,, \alpha=0.5\,, \mathrm{Pm}=1$. As the velocity perturbation is almost unchanged for different $\vec{B}_0$ for this Le, only one case is shown. $(b-h)$ represents the magnetic perturbations $\vec{B}$ with a Malkus field, Prendergast field, aligned dipolar field, tilted dipolar ($\theta_0=\pi/4$), tilted dipolar ($\theta_0=\pi/2$), free-decay poloidal dipole field and quadrupole fields, respectively. The black lines show field lines of the background magnetic field.}
       \label{fig:Iso-le1em3}
    \end{figure*}

    Fig.~\ref{fig:Iso-le1em3} presents pseudo-colour maps in the meridional plane and iso-surfaces within the shell for both the velocity field and magnetic perturbations at $\mathrm{Le}=10^{-3}$. At this relatively small Le, the velocity field is only weakly influenced by the background magnetic field, therefore only a single representative velocity field is shown. The flow is dominated by inertial waves, consistent with the structure shown in  
    Fig.~\ref{fig:zonalflow}(\textit{a}). The induced magnetic field exhibits distinct characteristics depending on the background magnetic configuration. For the Malkus field, the magnetic perturbation closely follows the structure of the velocity field, but its amplitude is significantly weaker than in other cases (as indicated by the colourbar). The Prendergast field produces a magnetic perturbation similar to that of an aligned dipolar field, although with a reduced amplitude, reflecting its comparable but weaker poloidal component. In general, the induced magnetic field is concentrated along inertial wave beams, especially those emanating from critical latitudes and in the polar regions, where both the background magnetic fields and flow velocity are strongest. For tilted dipolar fields, the induced magnetic field becomes asymmetric with respect to the equatorial plane, although it remains aligned with inertial wave structures and is still concentrated near the polar regions of the background field. The free-decay poloidal dipole and quadrupole fields generate magnetic perturbations qualitatively similar to those of the radially decaying dipolar field.

\begin{figure*}
	   \includegraphics[width=2\columnwidth]{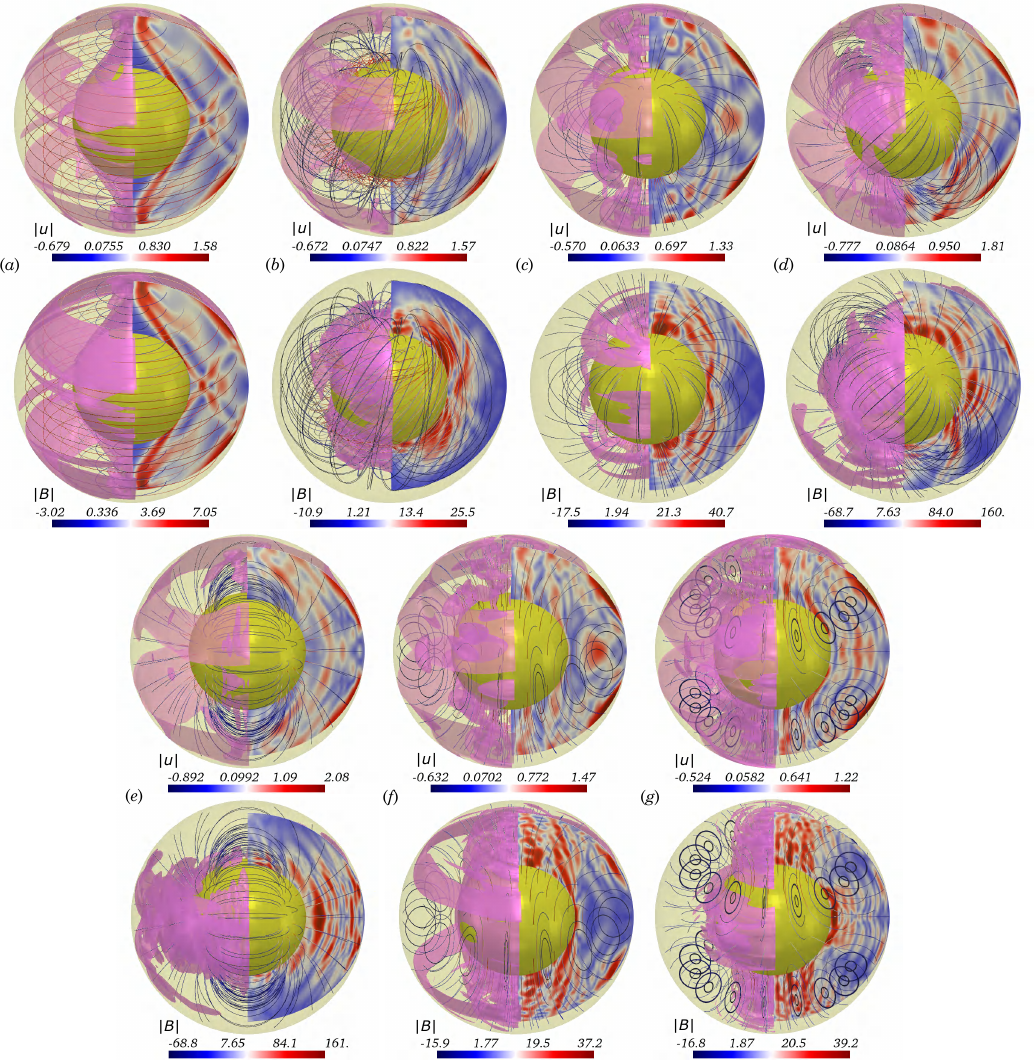}
       \caption{Pseudo-colour maps in the meridional plane and iso-surfaces in the shell for velocity $|\vec{u}|$ (upper) and magnetic perturbations $|\vec{B}|$ (bottom) when considering different background magnetic fields $\vec{B}_0$ at $\mathrm{Le}=10^{-2}$ and $\omega=1.1\,, \mathrm{Ek}=10^{-5}\,, \alpha=0.5\,, \mathrm{Pm}=1$.  $(a-g)$ represent the velocity perturbations $\vec{u}$ with the Malkus field, Prendergast field, aligned dipolar field, tilted dipolar ($\theta_0=\pi/4$), tilted dipolar ($\theta_0=\pi/2$), free-decay dipole field and quadrupole field, respectively. The black lines show field lines of the background magnetic field.}
       \label{fig:Iso-le1em2}
    \end{figure*}

\begin{figure*}
	   \includegraphics[width=2\columnwidth]{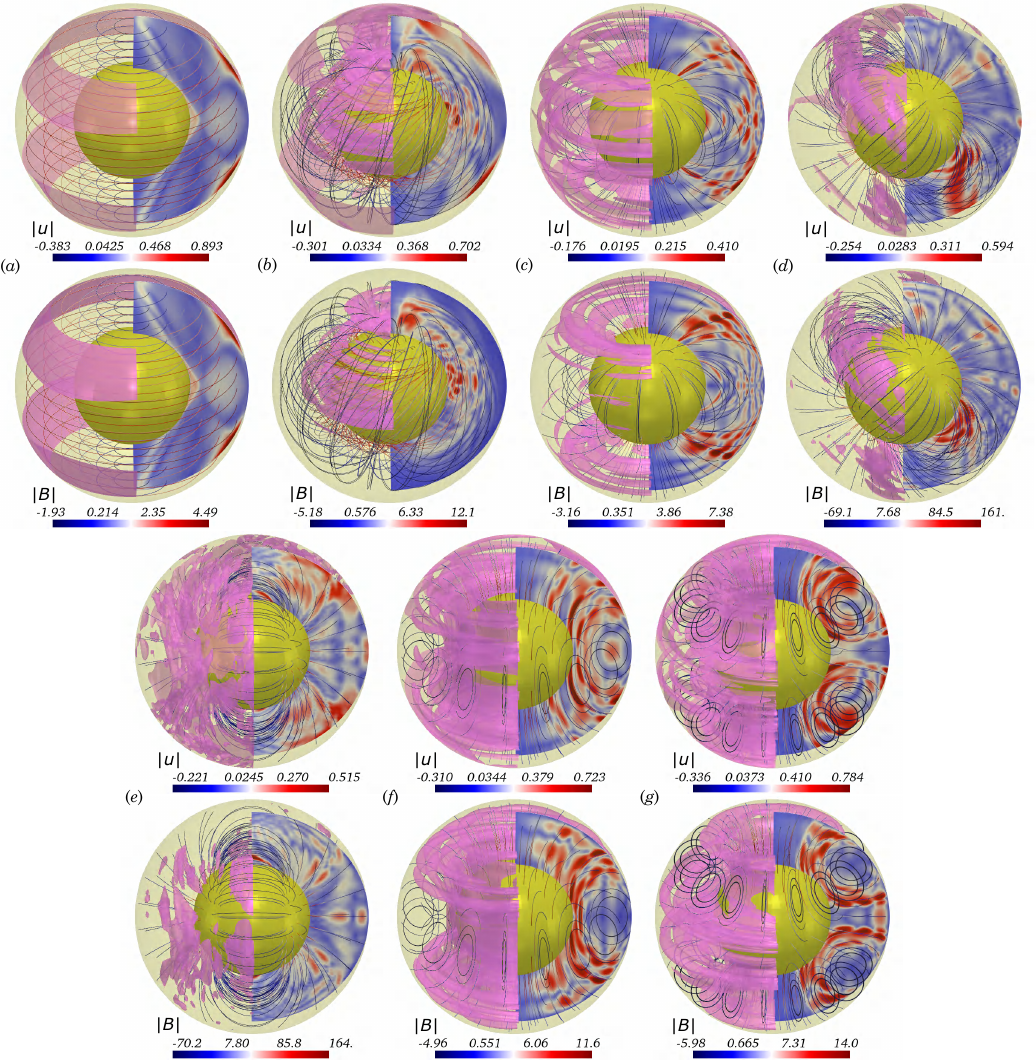}
       \caption{Pseudo-colour maps in the meridional plane and iso-surfaces in the shell for velocity $|\vec{u}|$ (upper) and magnetic perturbations $|\vec{B}|$ (bottom) when considering different background magnetic fields $\vec{B}_0$ at $\mathrm{Le}=5\times 10^{-2}$ and $\omega=1.1\,, \mathrm{Ek}=10^{-5}\,, \alpha=0.5\,, \mathrm{Pm}=1$. $(a-g)$ represent the velocity perturbations $\vec{u}$ with the  Malkus field, Prendergast field, aligned dipolar field, tilted dipolar ($\theta_0=\pi/4$), tilted dipolar ($\theta_0=\pi/2$), free-decay dipole field and quadrupole field, respectively. The black lines show field lines of the background magnetic field.}
       \label{fig:Iso-le5em2}
    \end{figure*}

    As we show in Fig.~\ref{fig:Iso-le1em2} for $\mathrm{Le}=10^{-2}$, inertial waves are largely suppressed by most background magnetic field configurations, with the notable exception of the Malkus field. In this case, both the velocity and magnetic perturbations remain nearly unchanged, indicating that a purely axisymmetric toroidal magnetic field has a relatively weak influence on the tidal flow  \citep[as also found by][for dissipation of flows excited by Earth's free inner core nutation]{LO2020}. For other magnetic configurations, the flow is predominantly governed by magneto-inertial waves, which arise from the combined action of the Coriolis and Lorentz forces. The tilted $\theta_0=\pi/4$ dipolar field breaks the symmetry of the tidal flow with respect to the equatorial plane, leading to clearly asymmetric structures. On the contrary, the tilted $\theta_0=\pi/2$ dipolar field preserves the equatorial symmetry of the tidal flow. For the poloidal free-decay dipole and quadrupole fields, the velocity perturbations tend to align with the rotation axis, while the induced magnetic perturbations are concentrated in the polar regions, where the background magnetic field is strongest. At $\mathrm{Le}=5\times 10^{-2}$ (Fig.~\ref{fig:Iso-le5em2}), inertial waves are suppressed for all background magnetic field configurations. Both the velocity and magnetic perturbations behave similarly to pure Alfvén waves, with magnetic tension acting as the dominant restoring force. In this regime, perturbations are strongly localised along magnetic field lines for all configurations. Resonant amplification will occur when the tidal forcing frequency matches the natural Alfvén frequency associated with individual field line. This will be shown in Figure ~\ref{fig:pt-eigen} in the next section. 
   
 In general, the evolution of the wave structure with increasing
    $\mathrm{Le}$ is qualitatively consistent with that reported by \citet{lin2018tidal}. They found that sufficiently weak magnetic fields
    leave the inertial-wave characteristics and wave attractors largely
    unchanged, whereas increasing $\mathrm{Le}$ causes the Lorentz force to
    modify and eventually disrupt these structures. At larger $\mathrm{Le}$,
    the response becomes increasingly Alfvénic and is organised along magnetic
    field lines. We observe the same broad transition here, from inertial-wave
    beams at $\mathrm{Le}=10^{-3}$, through more complex magneto-inertial waves
    at $\mathrm{Le}=10^{-2}$, to predominantly Alfvén-like propagation at
    $\mathrm{Le}=5\times10^{-2}$. However, unlike the axisymmetric fields
    considered by \citet{lin2018tidal}, the tilted dipolar field couples
    different azimuthal harmonics and thereby generates the oscillatory $m=0$
    zonal-flow component discussed above. The precise values of $\mathrm{Le}$
    at which these transitions occur also depend on the field geometry and
    dissipative parameters. Furthermore, their much smaller value of
    $\mathrm{Pm}=10^{-4}$, compared with $\mathrm{Pm}=1$ here, led to
    predominantly Ohmic rather than comparable viscous and Ohmic dissipation
    once magnetic effects became important.
    
 \subsection{Tidal dissipation as a function of frequency for various field strengths and configurations}
 \label{Pt_w}
 
\begin{figure*}
	   \includegraphics[width=2\columnwidth]{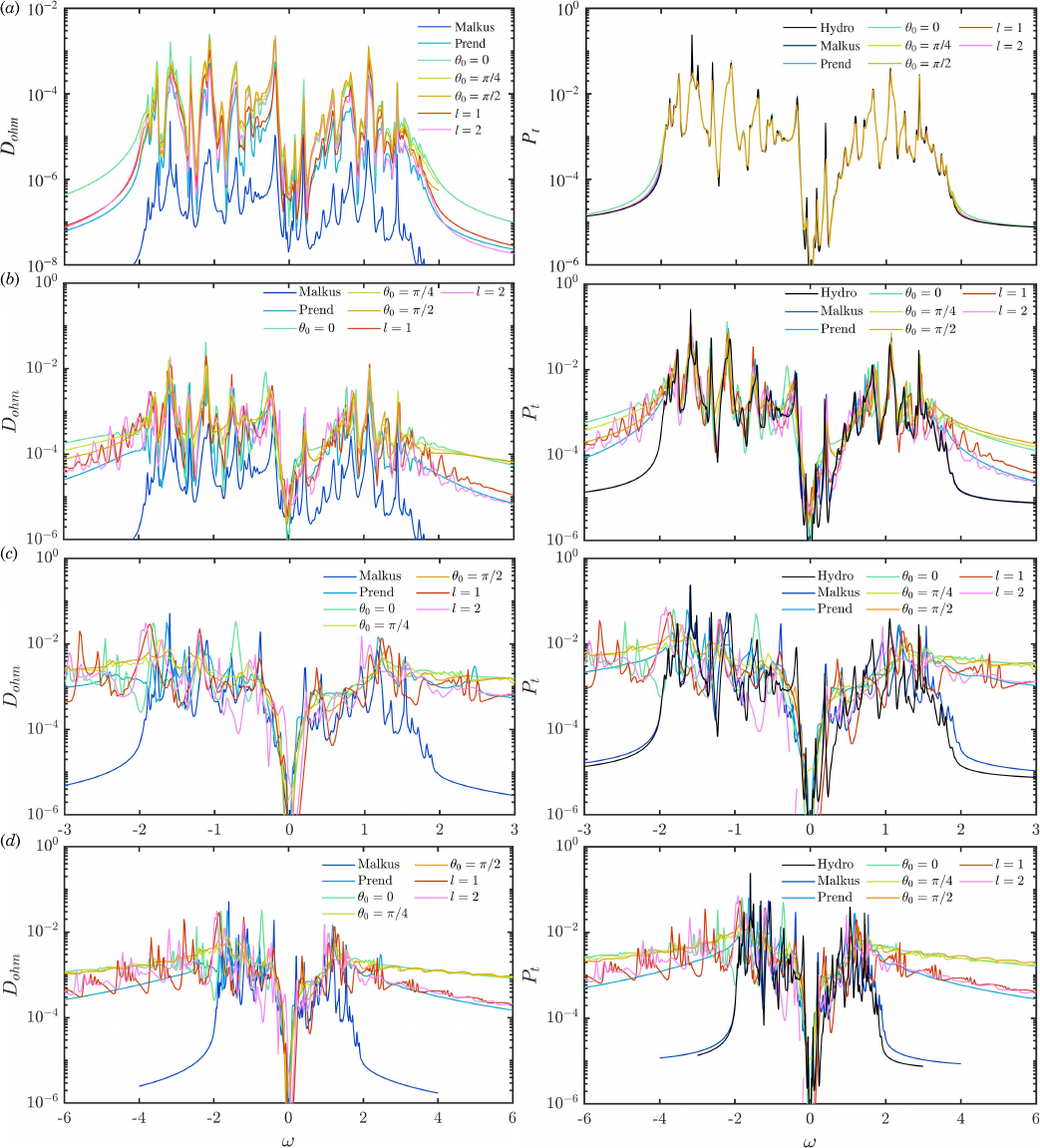}
       \caption{Ohmic dissipation $D_{\mathrm{ohm}}$ (left) and tidal power $P_\mathrm{t}$ (right) versus tidal frequency considering different background magnetic fields at $\alpha=0.5,\,\mathrm{Pm}=1,\, \mathrm{Ek}=10^{-5}$.  $(a)$ $\mathrm{Le}=10^{-3}$. $(b)$ $\mathrm{Le}=10^{-2}$. $(c)$ $\mathrm{Le}=5\times 10^{-2}$ for $-3<\omega<3$. $(d)$ $\mathrm{Le}=5\times 10^{-2}$ for $-6<\omega<6$.}
       \label{fig:PtDohm-le}
    \end{figure*}

    The Ohmic dissipation and tidal power as a function of frequency are shown in Fig.~\ref{fig:PtDohm-le} for various field strengths and configurations. These calculations were performed with a frequency resolution of $\Delta \omega=0.01$ for $\mathrm{Le}=10^{-3}$, except for tilted dipolar fields ($ \Delta \omega=0.02$) due to the higher computational cost of solving the associated initial value problem. For $\mathrm{Le}=10^{-2}$ and $\mathrm{Le}=5\times 10^{-2}$, a resolution of $ \Delta \omega=0.02$ was adopted. This was found to be sufficient in all cases to capture the relevant features (and for sufficiently accurate integrations over frequency to determine the frequency-averaged dissipation).
    
    At $\mathrm{Le}=10^{-3}$ (Fig.~\ref{fig:PtDohm-le}(\textit{a})), Ohmic dissipation is much smaller than the viscous dissipation, and its frequency dependence is nearly identical across all background magnetic fields, except for the Malkus field. This reflects the fact that the tidal flow is dominated by inertial waves and is only weakly influenced by the magnetic field at this $\mathrm{Le}$. 
    Accordingly, the tidal power remains in good agreement with the hydrodynamical result for all field configurations (for this Ek and Pm).
    
    At $\mathrm{Le}=10^{-2}$ (Fig.~\ref{fig:PtDohm-le}(\textit{b})), Ohmic dissipation becomes comparable to both viscous dissipation (not shown here) and tidal power. In the presence of such a strong magnetic field, the tidal response extends to higher forcing frequency magnitudes, outside the inertial range  ($-2<\omega<2$), consistent with the findings of \citet{lin2018tidal}. This is because the waves are restored by both Coriolis forces and magnetic tension, and fast magneto-inertial waves can propagate outside the range $-2<\omega<2$ for sufficiently large Le. For the Malkus field, however, the Ohmic dissipation remains relatively weak, and the tidal power closely follows the hydrodynamical case, further suggesting that purely axisymmetric toroidal fields have a limited impact on the linear tidal flow. 
    
    Within the inertial range ($-2<\omega<2$), the distributions of Ohmic dissipation and tidal power are broadly similar across all magnetic configurations (again excluding the Malkus field), with peak structures comparable to the hydrodynamical case. This indicates that inertial waves still dominate the dissipation in this regime. For $\omega>2$, the Ohmic dissipation and tidal power associated with the Prendergast field and the free-decay poloidal dipole and quadrupole fields are of similar magnitude, while dipolar and tilted dipolar fields exhibit comparable tidal power. Notably, multiple distinct peaks appear at specific frequencies for the free-decay dipole and quadrupole fields, likely associated with resonances that occur when the tidal frequency matches the natural frequencies of fast magneto-inertial oscillations. For large frequency magnitudes, these will resemble Alfv\'{e}n waves propagating along magnetic field lines.
    
At $\mathrm{Le}=5\times 10^{-2}$ (Fig.~\ref{fig:PtDohm-le}(\textit{c,d})), the presence of these magnetic fields significantly smooths out the tidal power spectrum, in agreement with the findings of \citet{lin2018tidal} for uniform axial and aligned dipolar magnetic fields. The Malkus field produces appreciable Ohmic dissipation within the inertial range  ($-2<\omega<2$), but still has a relatively weak influence on the overall tidal power distribution, reinforcing our conclusion that purely axisymmetric toroidal magnetic components are less effective in modifying linear $m=2$ tidal flows than fields with substantial poloidal components. In this regime, both Ohmic dissipation and tidal power maintain moderately large values to much higher frequencies than in the hydrodynamic case (reaching $\omega > 10$, not shown), particularly for dipolar and tilted dipolar fields, which exhibit a slow decay of tidal power with increasing frequency. Resonant features are evident across all dipolar-type fields, especially for tilted dipolar fields and free-decay poloidal dipole and quadrupole configurations.

\begin{figure}
	   \includegraphics[width=\columnwidth]{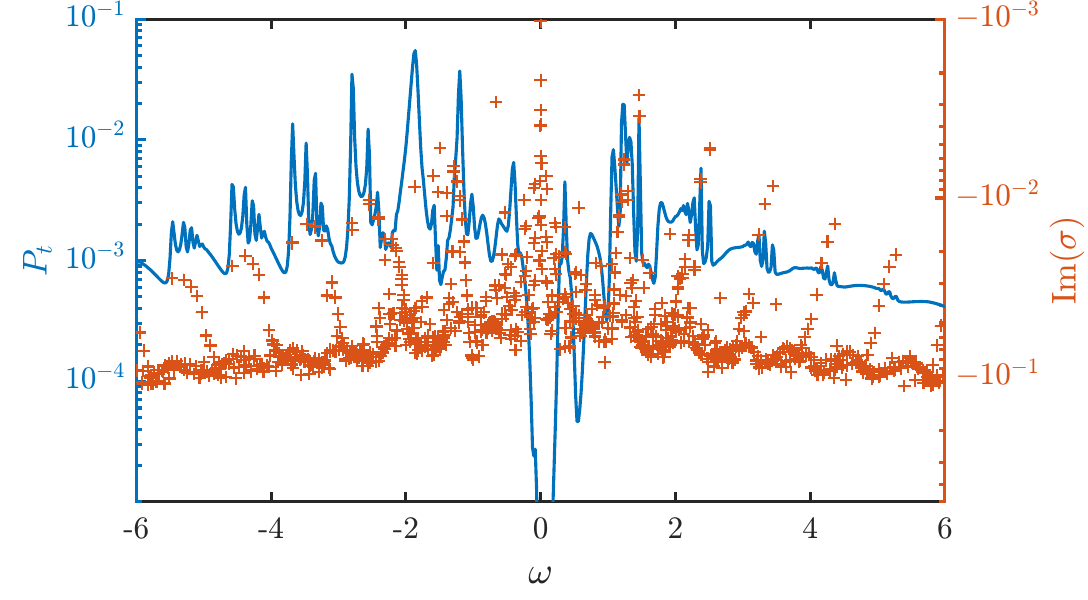}
       \caption{Tidal power versus tidal frequency together with the damping rates (imaginary parts of the frequency) and frequencies of the least-damped eigenmodes at $\mathrm{Le}=5\times 10^{-2}$ when a free-decay poloidal dipole field ($l=1$) is adopted as the background magnetic field.
       }
       \label{fig:pt-eigen}
    \end{figure}

    To validate our interpretation that the enhanced dissipation outside the inertial range shown in Fig.~\ref{fig:PtDohm-le} is associated with resonances with global magneto-inertial modes, we have solved the eigenvalue problem associated with equations ~(\ref{eq:mom}--\ref{eq:Bcon}). We show the resulting eigenvalues for a free-decay poloidal dipole field ($l=1$) as the background magnetic field in Fig.~\ref{fig:pt-eigen} (without tidal forcing), together with the tidal power as a function of forcing frequency. Here, $\omega$ denotes the eigenfrequency, while $\mathrm{Im}(\sigma)$ represents the damping rate of each mode (i.e., the imaginary part of the complex frequency). The distribution of peaks in the dissipation beyond the inertial range ($\left | \omega \right|>2$) agrees very well with the frequencies of the least-damped eigenmodes. This correspondence demonstrates that the observed peaks in tidal power arise from resonances between the tidal forcing frequency and the natural frequencies of weakly damped (fast) magneto-inertial eigenmodes. This provides direct evidence that the high-frequency tidal response is governed by resonances with the least-damped eigenmodes discrete (fast) magneto-inertial modes (which have a mostly Alfvénic character for the largest frequencies considered). On the other hand, while some dissipative peaks in the inertial range appear to be associated with resonances with certain least-damped eigenmodes, the picture is less clear-cut than for larger frequency magnitudes, perhaps due to the dense spectrum of waves in this frequency range, or because a one-to-one identification with individual inertial eigenmodes is hindered by the dense and partly singular modal structure, including responses associated with characteristics or wave attractors. 
    Further details of the tidal response outside the inertial range, and exploration of the magneto-inertial eigenmodes, will be presented in a forthcoming study.

  \subsection{Tidal response in shells with different convective envelope thicknesses $\alpha$} 
  
\begin{figure}
	   \includegraphics[width=\columnwidth]{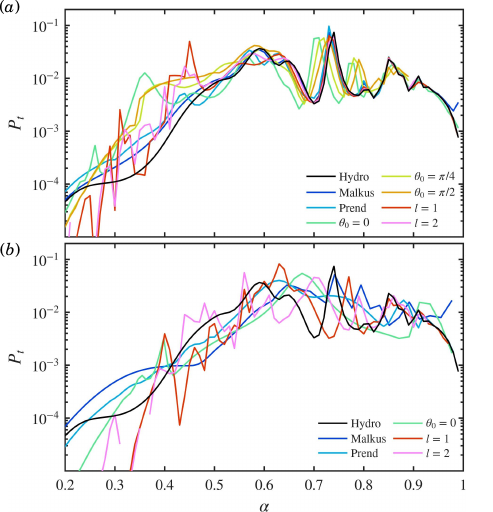}
       \caption{Tidal power $P_\mathrm{t}$ versus $\alpha$ considering various background magnetic fields for $\omega=1.1,\,\mathrm{Ek}=10^{-5},\,\mathrm{Pm}=1 $. $(a)$ $\mathrm{Le}=10^{-2}$;$(b)$ $\mathrm{Le}=5\times10^{-2}$. The tilted dipolar field is only considered at $\mathrm{Le}=10^{-2}$ due to the larger computational cost of exploring cases with $\mathrm{Le}=5\times 10^{-2}$ as an initial value problem.}
       \label{fig:pt-alpha}
    \end{figure}

    We have also investigated the effect of the shell thickness, characterised by $\alpha$, on the tidal response. Fig.~\ref{fig:pt-alpha} shows the variation of tidal power with $\alpha$ at a fixed frequency $\omega=1.1$ for $\mathrm{Le}=10^{-2}$ (panel (a)) and $\mathrm{Le}=5\times 10^{-2}$ (panel (b)). The Ohmic dissipation is not shown as it exhibits a distribution very similar to that of the tidal power. In addition, values for $\alpha < 0.2$ are omitted, since all quantities decay to negligible levels in this regime, as discussed in \S~\ref{Model}.
   In general, a thinner shell (larger $\alpha$) weakens the influence of the background magnetic field. 
   This trend is particularly evident in Fig.~\ref{fig:pt-alpha}(\textit{a}) for $\mathrm{Le}=10^{-2}$, where the tidal power for different magnetic configurations tends to converge towards the hydrodynamical value as $\alpha$ increases.  In contrast, for thicker shells (smaller $\alpha$), the magnetic field plays a more significant role, leading to substantial differences in tidal power among different field configurations. At $\mathrm{Le}=5\times 10^{-2}$, although the tidal power does not fully converge to the hydrodynamical value, the variation amongst different magnetic field cases decreases as $\alpha\to 1$. This behaviour is primarily attributed to the strengthening of inertial wave excitation in thinner shells \citep[][]{lin2018tidal}, potentially making them more resistant to magnetic modification. Consequently, a stronger magnetic field is required to significantly alter the tidal flow in this regime. This highlights the competing roles of geometric confinement and magnetic tension in controlling the tidal response in rotating and magnetised fluid envelopes. 

\begin{figure}
	   \includegraphics[width=\columnwidth]{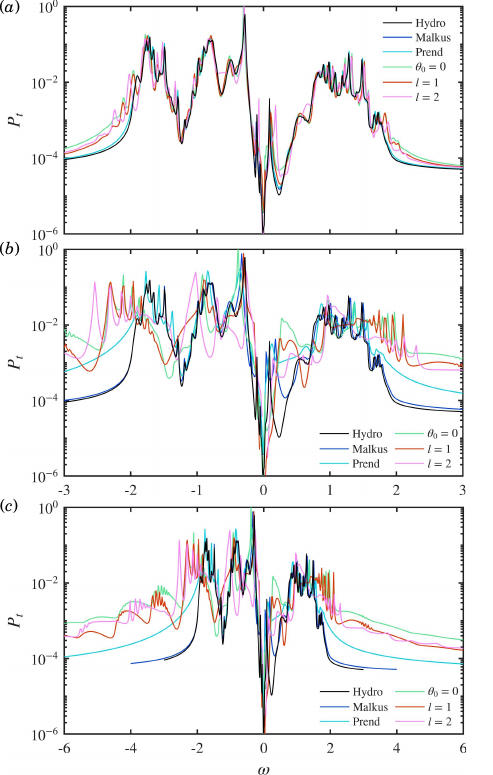}
       \caption{Tidal power $P_{\mathrm{t}}$ versus tidal frequency considering different background magnetic field strengths and configurations for $\alpha=0.8,\, \mathrm{Pm}=1\,, \mathrm{Ek}=10^{-5}$; $(a)$ $\mathrm{Le}=10^{-2}$. $(b,c)$ $\mathrm{Le}=5\times 10^{-2}$ with $-3<\omega<3$ and $-6<\omega<6$.}
       \label{fig:pt-alpha-om}
    \end{figure}

    The frequency dependence of the tidal power for a thinner shell ($\alpha=0.8$) is shown in Fig.~\ref{fig:pt-alpha-om} at $\mathrm{Le}=10^{-2}$ and $\mathrm{Le}=5\times 10^{-2}$. Owing to the more substantial computational cost associated with tilted dipolar fields, these cases are not included here and in the subsequent analysis. At $\mathrm{Le}=10^{-2}$, the tidal power for different background magnetic field configurations is in close agreement with the hydrodynamical case. This contrasts with the behaviour observed for a thicker shell ($\alpha=0.5$, see Fig.~\ref{fig:PtDohm-le}(\textit{b})), and further supports the conclusion that a thinner shell reduces the influence of the background magnetic field (at least for the parameters we have considered). At $\mathrm{Le}=5\times 10^{-2}$, the tidal power associated with the Malkus field remains nearly identical to the hydrodynamical result, while other magnetic field configurations produce distinct tidal power distributions. Multiple resonance peaks are still present at higher frequencies, although they are less regular than those observed in the thicker shell ($\alpha=0.5$). In addition, the tidal power decays more rapidly at high frequencies, indicating that Alfvén-like wave activity is suppressed in thinner shells. This suggests that geometric confinement not only weakens magnetic coupling but also disrupts the coherence of the excitable Alfvénic resonances at high frequencies. 

  \subsection{Tidal response for various $\mathrm{Pm}$ and $\mathrm{Ek}$}

In this section, we explore the effects of varying the diffusivities on the wavelike tidal response, and in particular the values of the Ekman number $\mathrm{Ek}$ and magnetic Prandtl number $\mathrm{Pm}$. This is important whether or not we use turbulent values (in which cases the correct values of Ek and Pm are uncertain) or microscopic ones (in which case we would need to extrapolate to astrophysical parameters as realistic values are unachievable). In particular, turbulent values of Pm are likely to be $O(1)$ whereas microscopic values are typically $Pm\approx 10^{-2}$ in stellar convective envelopes but could take even smaller values in giant planets. 
  
\begin{figure*}
	   \includegraphics[width=2\columnwidth]{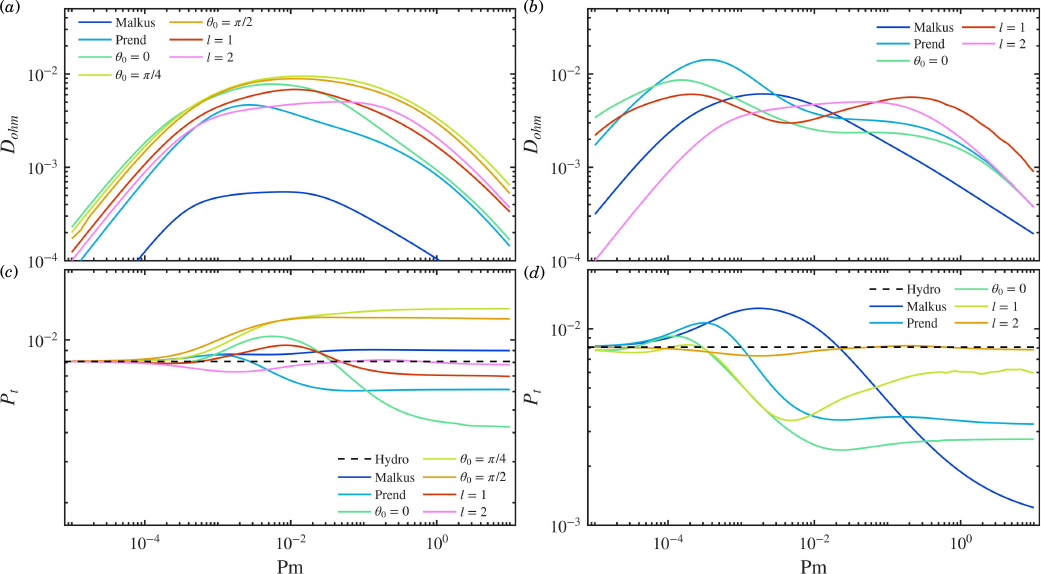}
       \caption{Ohmic dissipation $D_{\mathrm{ohm}}$ versus $\mathrm{Pm}$ considering various background magnetic fields at $(a)$ $\mathrm{Le}=10^{-2}$  and $(b)$ $\mathrm{Le}=5\times 10^{-2}$. Tidal power $P_\mathrm{t}$ versus $\mathrm{Pm}$ considering various background magnetic fields at $(c)$ $\mathrm{Le}=10^{-2}$ and $(d)$ $\mathrm{Le}=5\times 10^{-2}$. The parameters are $\omega=1.1,\,\alpha=0.5,\, \mathrm{Ek}=10^{-5} $. The tilted dipolar field is only considered at $\mathrm{Le}=10^{-2}$ due to the greater computational costs of performing the initial value problem at $\mathrm{Le}=5\times 10^{-2}$.}
       \label{fig:pt-pm}
    \end{figure*}

We vary Pm over the range from $10^{-5}$ to $10$, for fixed tidal frequency $\omega=1.1$, while fixing $\mathrm{Ek}=10^{-5}$, in order to assess its influence on the tidal power. The Ohmic dissipation and tidal power as functions of $\mathrm{Pm}$ are shown in Fig.~\ref{fig:pt-pm}. Very small and very large values of $\mathrm{Pm}$ both lead to reduced Ohmic dissipation for $\mathrm{Le}=10^{-2}$ and $\mathrm{Le}=5\times10^{-2}$. The maximum Ohmic dissipation occurs at intermediate values, around $\mathrm{Pm}\approx 10^{-2}$ and $\mathrm{Pm}\approx 10^{-3}$ for $\mathrm{Le}=10^{-2}$ and $\mathrm{Le}=5\times10^{-2}$, respectively. In the limit of very small $\mathrm{Pm}$, the tidal power converges towards the hydrodynamical value, owing to the strong Ohmic diffusion acting on magnetic perturbations. As $\mathrm{Pm}$ is increased, the tidal power approaches a nearly constant value, indicating that the tidal flow becomes insensitive to further increases in $\mathrm{Pm}$. The reduction in Ohmic dissipation at the largest $\mathrm{Pm}$ here can therefore be attributed to the decrease in magnetic diffusivity (i.e., smaller $\mathrm{Em}$), while the underlying flow remains essentially unchanged. This suggests that $\mathrm{Pm}$ primarily controls the relative efficiency of magnetic diffusion rather than the overall response of the tidal flow itself for the parameters considered here. 

\begin{figure}
	   \includegraphics[width=\columnwidth]{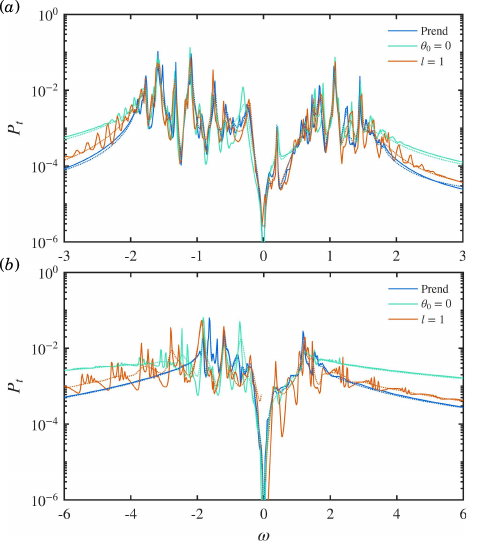}
       \caption{Tidal power $P_{\mathrm{t}}$ versus tidal frequency considering different background magnetic fields at $\alpha=0.5\,,\mathrm{Ek}=10^{-5}$; $(a)$ $\mathrm{Le}=10^{-2}$ and $(b)$ $\mathrm{Le}=5\times 10^{-2}$. Solid lines represent $\mathrm{Pm}=1$ while dotted lines mean the cases at $\mathrm{Pm}=10^{-2}$.
       }
       \label{fig:pt-pm-om}
    \end{figure}

A broad frequency scan of the tidal power as a function of frequency further supports this interpretation. Fig.~\ref{fig:pt-pm-om} compares the tidal power for $\mathrm{Pm}=1$ (solid lines) and $\mathrm{Pm}=10^{-2}$ (dotted lines) for $\mathrm{Ek}=10^{-5}$ and $\alpha=0.5$. For clarity, only three representative background magnetic field configurations are shown. Overall, the tidal power for $\mathrm{Pm}=10^{-2}$ and $\mathrm{Pm}=1$ agrees well, particularly within the inertial range ($-2<\omega<2$). At higher frequencies, where Alfvénic wave resonances dominate, the resonant peaks are noticeably suppressed for $\mathrm{Pm}=10^{-2}$ due to the enhanced Ohmic diffusivity in this case. This behaviour is observed at both $\mathrm{Le}= 10^{-2}$ and $\mathrm{Le}=5\times 10^{-2}$. Similar trends are found for other magnetic field configurations. This confirms that magnetic diffusivity primarily damps high-frequency Alfvénic resonances, while leaving the inertial-wave-dominated response in the inertial range largely unaffected for the parameters we have considered. 
    
\begin{figure}
	   \includegraphics[width=\columnwidth]{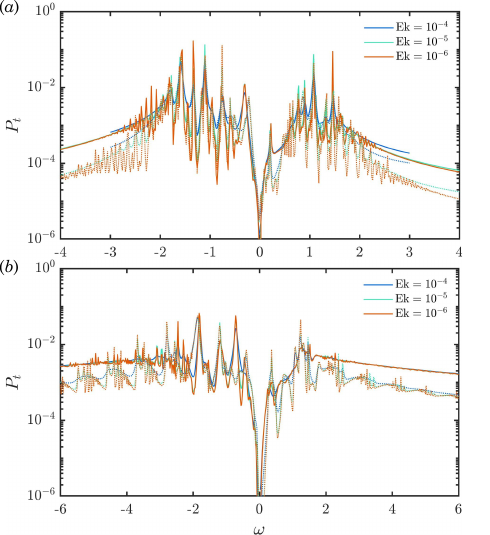}
       \caption{Tidal power $P_{\mathrm{t}}$ versus tidal frequency considering different background magnetic fields at $\alpha=0.5$ and different $\mathrm{Ek}$; $(a)$ $\mathrm{Le}=10^{-2}$ and $(b)$ $\mathrm{Le}=5\times 10^{-2}$. Solid lines represent cases with a (radially decaying) dipolar field, while dotted lines represent cases with a free-decay poloidal dipole field.
       }
       \label{fig:pt-ek-om}
    \end{figure}

    We examine the effects of varying $\mathrm{Ek}$ by considering three values ($10^{-4}, 10^{-5}, 10^{-6}$) at $\mathrm{Le}=10^{-2}$, $\mathrm{Le}=5\times10^{-2}$ and fixing $\mathrm{Pm}=1$ . For comparison, we focus on the dipolar field and the free-decay poloidal dipole field, as shown in Fig.~\ref{fig:pt-ek-om}. At $\mathrm{Le}=10^{-2}$, decreasing $\mathrm{Ek}$ does not lead to significant changes in the tidal power within the inertial range for either magnetic configuration. The locations and amplitudes of the peaks remain in good agreement across all three values of $\mathrm{Ek}$. The main differences arise in the Alfvén-wave-dominated regime ($\omega > 2$), where there are more resonance peaks as $\mathrm{Ek}$ is decreased. This effect is particularly pronounced for the free-decay poloidal dipole field. At $\mathrm{Le}=5\times10^{-2}$, similar trends are observed. The Alfvénic resonance peaks become increasingly prominent at smaller $\mathrm{Ek}$, especially at $\mathrm{Ek}=10^{-6}$ for both magnetic field configurations. However, for the dipolar field, resonant features remain less pronounced at $\omega > 2$ for both $\mathrm{Le}=10^{-2}$ and $\mathrm{Le}=5\times10^{-2}$. Overall, reducing $\mathrm{Ek}$ enhances the visibility of Alfvénic resonances while having only a minor impact on inertial-wave-dominated dynamics. This reflects the reduced viscous damping at smaller $\mathrm{Ek}$, which allows weakly damped Alfvénic modes to be more clearly excited.

    We note that the ranges of $\mathrm{Ek}$ and $\mathrm{Pm}$ explored here are relatively limited by the computational costs of exploring very small Ek and Pm, and therefore our results do not aim to provide a comprehensive parameter survey. Instead, they are intended to illustrate the most important changes in the roles of viscous and magnetic diffusion in shaping the tidal response. In particular, the trends identified here highlight how $\mathrm{Ek}$ and $\mathrm{Pm}$ primarily influence the Alfvénic component of the flow, though the quantitative behaviour may differ in other parameter regimes.
    
  \subsection{Frequency-averaged tidal dissipation}
  
     The frequency-averaged dissipation rate or tidal power is a useful way to quantify the overall dependence of inertial wave dissipation on the structure of a body without focusing on particular details of the frequency-dependent response \citep{ogilvie2013tides}. In addition, because there is a certain frequency-weighted average that is straightforward and computationally inexpensive to compute -- using an analytical expression for a piece-wise homogeneous body, or from solving an ODE for more realistic models, rather than solving two-dimensional boundary value problems or performing direct numerical simulations -- this has provided a useful way to explore the long-term evolution of the spin and orbital parameters in stellar and planetary systems through tidal angular momentum exchanges caused by inertial waves \citep[e.g.][]{BM2016,bolmont2017water,gallet2017tidal,B2020,Barker2022}. One advantage of this measure is that it is independent of the dissipative properties of the fluid. Interestingly, \citet{lin2018tidal} also found that the frequency-averaged dissipation obtained in their linear calculations with an aligned dipole magnetic field and uniform axial field were remarkably consistent with hydrodynamical predictions, and they provided arguments to justify this observation. Here, we aim to test whether this conclusion holds for the different background magnetic fields that we have been exploring in this paper.

     The particular frequency-averaged tidal power that is most commonly considered is defined by the dimensionless quantity
     \begin{align}
       \Lambda =\int_{-\infty}^{\infty }\mathrm{Im}\left [ k_{l}^{m}(\omega) \right ]\frac{\mathrm{d}\omega}{\omega},
      \label{eq:fre-average}  
      \end{align}
      where $\mathrm{Im}\left [ k_{l}^{m} \right ]$ is the imaginary part of the potential Love number.  For the tidal component with $l=m=2$,  $\mathrm{Im}\left [ k_{2}^{2}(\omega) \right ]$ is related to the tidal power by 
      \begin{align}
       \left | \mathrm{Im}\left [ k_{2}^{2}(\omega) \right ]\right |=\frac{6}{5}\frac{\epsilon_t^2}{C_t^2}\frac{P_\mathrm{t}(\omega)}{\left | \omega\right |}.
       \label{eq:fre-love}  
       \end{align}
        Here, $\epsilon_t=(1+k_2)\epsilon_\Omega$, and $\epsilon^2_\Omega=\Omega^2R^3/GM$ is a small parameter for astrophysical bodies and is usually not well-known unless the rotation period has been constrained.

        In an incompressible fluid body containing a rigid core, \citet{ogilvie2013tides} derived an analytical expression for $\Lambda$ by considering the low-frequency hydrodynamic response to an impulsive forcing, which can be written
        \begin{align}
               \Lambda=\frac{16\pi}{63}\epsilon _t^2\left(\frac{\alpha^5}{1-\alpha^5}\right).
               \label{eq:fre-theo}  
        \end{align}
         By combining equations ~(\ref{eq:fre-average},~\ref{eq:fre-love},~\ref{eq:fre-theo}), we obtain
         \begin{align}
              \int_{-\infty }^{\infty }\frac{P_\mathrm{t}(\omega)}{\omega^2 C_t^2}\mathrm{d}\omega=\frac{40\pi}{189}\left(\frac{\alpha^5}{1-\alpha^5}\right).
               \label{eq:fre-final}  
        \end{align}
        The values of the frequency-averaged tidal power in our calculations, and the corresponding predicted value according to Eq.~(\ref{eq:fre-final}), are presented in table~\ref{tab:Pt_f}. 
        
        Most of the simulated frequency-averaged tidal power values are in good agreement with the theoretically predicted values. There are small differences for the cases with tilted dipolar magnetic fields at $\alpha=0.5, \mathrm{Pm}=1, \mathrm{Ek}=10^{-5}$, in particular. This is almost certainly caused by the lower frequency resolution used for these (due to its computational expense), potentially causing some features in the tidal power to have been inaccurately represented. Nevertheless, these values still agree to within a few percent even for the smallest Ekman numbers considered.
        
        Notably, there is an obvious difference (up to $37\%$) between the numerically-computed and predicted values for the free-decay poloidal dipole and quadrupole fields at $\mathrm{Le}=5\times10^{-2}$. This discrepancy can be attributed to a breakdown of the assumptions underlying the impulsive approach in \citet{ogilvie2013tides}. In the standard theory, the tidal response is decomposed into a non-wavelike (equilibrium tide) component and a wavelike component, and the Coriolis force acting on the former provides a well-defined impulsive forcing for the latter. This construction implicitly assumes that the non-wavelike component remains quasi-hydrostatic and that the resulting impulsive velocity field is unaffected by the magnetic field, such that the work done by this Coriolis force provides an accurate representation of the energy injected into the wave-like subspace. However, in the presence of a sufficiently strong, spatially structured and non-curl-free poloidal magnetic field, these assumptions may no longer be valid. The background Lorentz force contributes directly to the large-scale force balance, so that the work done by the magnetic field can compete with the tidally-driven energy exchanges. [A separate point is whether the Lorentz force acting on the non-wavelike tide should be considered also \citep[which was ignored in][]{lin2018tidal} \citep[see also][]{Astoul2019}, which we will defer to future work.]
        
\begin{figure}
	       \includegraphics[width=\columnwidth]{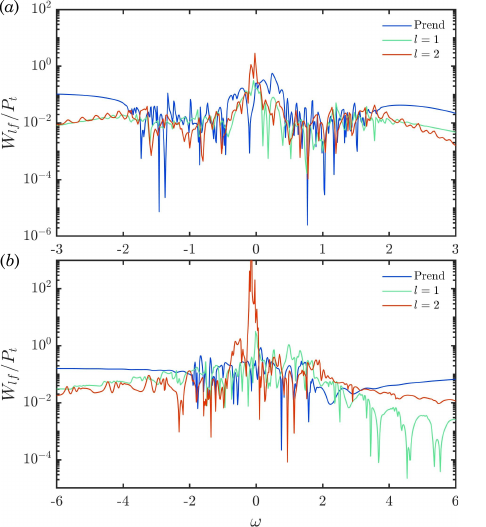}
          \caption{Ratio of work done by background Lorentz forces $W_{\mathrm{lf}}$ to the tidal power $P_\mathrm{t}$ for cases with $\mathrm{Ek}=10^{-5},\alpha=0.5,\mathrm{Ek}=10^{-5}$. $(\textit{a})$ $\mathrm{Le}=10^{-2}$;$(\textit{b})$ $\mathrm{Le}=5\times 10^{-2}$.
        }
        \label{fig:pt_wfl}
        \end{figure}

        Figure~\ref{fig:pt_wfl} shows the ratio of the work done by the background Lorentz force, $W_{\mathrm{lf}}$, to the tidal power input, $P_\mathrm{t}$, for the three background magnetic fields which have non-zero curl (current). The contribution of the background Lorentz force is non-negligible within the inertial frequency range, particularly for the free-decay poloidal dipole and quadrupole fields at $\mathrm{Le}=5\times10^{-2}$.
        In particular, $\int_{-\infty }^{\infty }\frac{W_{\mathrm{lf}}(\omega)}{\omega^2}\mathrm{d}\omega= 0.0030,0.0087$ for $l=1,2$, respectively, which is approximately consistent with the discrepancy between the simulated values and the theoretical hydrodynamic prediction for the frequency-averaged tidal power (note that this does not hold for the Prendergast field). This indicates that energy transfers mediated by the background field can play a significant role in modifying the tidal response in these cases, thereby potentially rendering the theory for the frequency-averaged dissipation inapplicable.

For astrophysical applications, we can crudely estimate the ratio,
\begin{align}
    \frac{W_\mathrm{lf}}{P_\mathrm{t}}\sim \frac{|\boldsymbol{u}\cdot(\nabla\times\boldsymbol{B}_0)\times \boldsymbol{b}|/\mu_0}{|\rho \boldsymbol{f}_\mathrm{t}|}\sim \frac{\omega_A^2}{\omega\Omega}\frac{R}{\ell},
    \label{Wlfratio}
\end{align}
after reintroducing dimensions throughout, if $b\sim (\omega_A/\omega)u\sqrt{\rho\mu_0}(R/\ell)$ 
from the induction equation, $\omega_A\sim B_0/(R\sqrt{\mu_0\rho})$, $u\sim C_t R \omega$, $\ell$ is a length-scale for the waves and we take $\nabla \times \boldsymbol{B}_0\sim B_0/R$. If we take $B_0\sim 1$ kG as the large-scale mean solar magnetic field in the convection zone, and adopt the volumetric mean density of the solar convection zone \citep[from the models in e.g.][]{JCD1996}, $\rho\sim 54 \mathrm{kg}/\mathrm{m}^3$ 
and take $\ell\sim R$ and the solar radius for $R$, then $\omega_A\sim 2\times10^{-8} \mathrm{s}^{-1}$. For a hot Jupiter on a one-day orbit around a slowly-rotating star, $\omega\sim 1.5\times 10^{-4} \mathrm{s}^{-1}$. 
For a 10-day stellar rotation period, this gives $W_\mathrm{lf}/P_\mathrm{t}\sim 3\times 10^{-7}$. Hence, the work done by Lorentz forces is likely to be much smaller than the tidal work in applications according to this estimate.

A significant caveat here is that the wavelike velocity amplitude for linear inertial waves (for $\mathrm{Le}\ll 1$) may be larger than we assumed above for small $\mathrm{Ek}$, with $u\sim \mathrm{Ek}^{-\beta}R\omega C_t$ with $\beta\approx 1/6-1/5$ \citep[e.g.][]{astoul2022effects}. Similarly, the lengthscale $\ell$ could be much smaller, where $\ell\sim \mathrm{Ek}^{\alpha_1}R$ with $\alpha_1\approx 1/4-2/5$ may be expected for linear inertial waves (for $\mathrm{Le}\ll 1$). These considerations would provide an additional factor of $\mathrm{Ek}^{-(\alpha_1+\beta)}$ on the right hand side of Eq.~\ref{Wlfratio}. Since $\mathrm{Ek}\ll 1$, this factor may substantially increase the ratio $W_\mathrm{lf}/P_\mathrm{t}$ for a given magnetic field strength, by factors of $10^3$ if $\mathrm{Ek}=10^{-5}$ (turbulent value), or up to $10^9$ if $\mathrm{Ek}=10^{-15}$ (microscopic value).

We can also estimate that the energy in the magnetic field $d^3 B_0^2/2\mu_0\sim 3.2\times 10^{28}$ J (where $d\sim 2\times 10^8$ m is the depth of the convection zone) is likely to be at most comparable to the total kinetic energy associated with the tidal flow $\rho d^3 C_t^2 R^2 \omega^2/2\sim  2.3\times 10^{28}$ J ($C_t\sim 10^{-4}$),
and much smaller than the kinetic (thermal much larger by 7 orders of magnitude!) energy in the convection, estimated to be $\gtrsim 10^{31}$ J \citep[e.g., using the numbers in][for convective velocities]{Birch2024}. Hence, the magnetic field is unlikely to be the dominant energy source.  However, if there are locally strong fields or currents, the magnetic energy could exceed the tidal energy and the work done by Lorentz forces could be non-negligible. For example, if $B_0\sim 100$ kG in the deep portions of the convection zone, and if the lengthscale of the field is $d\sim 2\times 10^8$ m, then $W_\mathrm{lf}/P_\mathrm{t}\sim 0.03$, and the magnetic energy would exceed the energy in the tidal flow or convection. Furthermore, it is possible that $W_\mathrm{lf}/P_\mathrm{t}$ could be very large on small scales, so the work done by Lorentz forces could be dynamically very important in practice for inertial waves even if the global energy balance arguments here suggest it to be subdominant.
        
        \begin{table*}
        \begin{center}
        \centering
        \caption{Frequency-averaged tidal power for all simulated cases. The frequency ranges used for the integrations are $\omega\in[-3,3]$ for $\mathrm{Le}=10^{-3}$ and $10^{-2}$, and $\omega\in[-8,8]$ for $\mathrm{Le}=5\times10^{-2}$. The frequency resolutions $\Delta\omega$ used for the cases at $\mathrm{Ek}=10^{-4}$ and $10^{-5}$ are specified at the beginning of Section~\ref{Pt_w}. For the corresponding cases at $\mathrm{Ek}=10^{-6}$, we use $\Delta\omega=0.005$ for the hydrodynamic and Malkus-field cases and $\Delta\omega=0.01$ for all other cases. These resolutions have been verified to ensure acceptable convergence of the frequency-averaged tidal power. }
        \label{tab:Pt_f}
        \begin{tabular}{c|cccccccccc}
        \hline
         Parameters & Le    & Predicted & Hydro  & Malkus & Prend  & $\theta_0=0$ & $\theta_0=\pi/4$ & $\theta_0=\pi/2$ &$l=1$ &$l=2$ \\ \hline
        \multirow{3}{*}{
                        $\alpha=0.5$, 
                          $\mathrm{Pm = 1}$,  
                        $\mathrm{Ek}=10^{-4}$
                        }    & 0.001 & 0.0214    & 0.0214 &0.0214  &0.0214  &0.0211      &-         &-  &0.0214 &0.0214 \\ 
                                              & 0.01  & 0.0214    & 0.0214 &0.0214  &0.0214  &0.0212      &-         &-  &0.0214  &0.0214 \\ 
                                              & 0.05  & 0.0214    & 0.0214 &0.0214  &0.0211  &0.0212      & -        &-  &0.0191  &0.0138      \\ \hline
        \multirow{3}{*}{
                        $\alpha=0.5$, 
                          $\mathrm{Pm} = 1$, 
                        $\mathrm{Ek}=10^{-5}$
                        }    & 0.001 & 0.0214    & 0.0214 & 0.0218 & 0.0217 & 0.0210     & 0.0195        & 0.0196 &0.0204 & 0.0204\\ 
                                              & 0.01  & 0.0214    & 0.0214 & 0.0218 & 0.0211 & 0.0214     & 0.0216        & 0.0218 &0.0224 &0.0220 \\ 
                                              & 0.05  & 0.0214    & 0.0214 & 0.0217 & 0.0214 & 0.0212     & 0.0210        &0.0213 &0.0188 &0.0138       \\ \hline
        \multirow{2}{*}{
                        $\alpha=0.8$, 
                          $\mathrm{Pm} = 1$, 
                        $\mathrm{Ek}=10^{-5}$
                        }
                        & 0.01  & 0.3241    & 0.3247 &0.3205  &0.3258 &0.3431      & -           & -   &0.3184 &0.3516 \\ 
                                              & 0.05  & 0.3241    & 0.3247 &0.3236  &0.3236  &0.3179      & -           & - &0.1933 &0.1251    \\ \hline
        \multirow{2}{*}{
                        $\alpha=0.5$, 
                        $\mathrm{Pm}=10^{-2}$, 
                        $\mathrm{Ek}=10^{-5}$
                        } & 0.01  &0.0214    & 0.0214 & 0.0210 & 0.0213 & 0.0213     & -           & -  &0.0215 &0.0214 \\ 
                                               & 0.05  &0.0214    & 0.0214 & 0.0214 & 0.0218 & 0.0213     & -           & -  &0.0205 &0.0136 \\ \hline
        \multirow{2}{*}{
                        $\alpha=0.5$, 
                        $\mathrm{Pm}=1$, 
                        $\mathrm{Ek}=10^{-6}$
                        } & 0.01  &0.0214    & 0.0216  & 0.0215  &-  &0.0223      & -           & -  &0.0242 &0.0216 \\ 
                                               & 0.05  & 0.0214    & 0.0216  & 0.0213  &- &0.0211      & -           & -  &0.0192 &0.0135 \\ \hline
        \end{tabular}
        \end{center}
        \end{table*}
        
\section{Conclusions}\label{conclusions}
   In this paper, we have performed linear magnetohydrodynamic calculations of tidally forced flows in rotating spherical shells to investigate how different magnetic field configurations modify tidal dissipation in the convective regions of stars and planets. Using a combination of boundary value calculations and direct numerical simulations, we have explored a broad range of magnetic field geometries, including aligned and tilted dipolar fields, a purely axisymmetric toroidal Malkus field, a mixed poloidal--toroidal Prendergast field, and free-decay axisymmetric poloidal dipole and quadrupole fields. We have also examined the dependence of the tidal response on the Lehnert number (strength of magnetic field to rotation), the tidal frequency, the shell thickness, the magnetic Prandtl number and the Ekman number.

   Our results show that the detailed tidal response at a given forcing frequency depends strongly on both the strength and the geometry of the background magnetic field. As the Lehnert number increases, the tidal flow undergoes a clear transition from an inertial-wave-dominated regime to a magnetically modified regime dominated by magneto-inertial waves, and for sufficiently large frequencies, to an Alfvénic regime when magnetic tension becomes sufficiently strong \citep[as also found by][in their calculations for a uniform axial and aligned dipolar field]{lin2018tidal}. A key result of this work is that the magnetic geometry plays an important role in determining how strongly the tidal response is modified. Background fields with substantial poloidal components, such as dipolar fields and the free-decay poloidal dipole and quadrupole fields, alter the tidal flow more significantly and at much smaller values of $\mathrm{Le}$ than the purely (axisymmetric) toroidal Malkus field \citep[as briefly reported for the problem of Earth's inner core nutation in][]{LO2020}. This indicates that the coupling between the tidal flow and the background field is much more effective when the field possesses a significant poloidal component. The Prendergast field produces intermediate behaviour, consistent with its mixed poloidal--toroidal character.
   Another important result concerns the distinction between axisymmetric and non-axisymmetric magnetic fields. For all axisymmetric configurations considered here (Malkus, Prendergast, aligned dipolar, and free-decay poloidal fields), the periodic tidal flow is steady. In contrast, tilted dipolar fields produce a time-periodic response, characterised by oscillations in tidal power and a clear anti-correlated exchange between kinetic and magnetic energies. They also generate oscillatory zonal flows even within the linear regime. This demonstrates that misalignment between the magnetic field and rotation axis can introduce qualitatively new tidal dynamics.
   
   We have also shown that magnetic fields can substantially widen the range of frequencies over which enhanced tidal dissipation occurs. In the hydrodynamical problem, the strongest response is confined to the inertial range, $|\omega|<2$, whereas sufficiently strong magnetic fields can extend the response to substantially higher frequencies. At intermediate and large $\mathrm{Le}$, the Ohmic dissipation becomes comparable to the viscous dissipation, and the tidal power remains significant well beyond the inertial range. At the same time, our calculations indicate that, for the parameters considered here ($\mathrm{Ek}\in[10^{-6},10^{-5}],\mathrm{Pm}=O(1)$), Ohmic dissipation does not necessarily dominate the total dissipation at large $\mathrm{Le}$, as reported in \citet{lin2018tidal}. Instead, (turbulent) viscous dissipation remains important, especially when a relatively large (compared to the microscopic value) effective Ekman number is adopted to represent turbulent viscosity (and associated Pm of order unity). This suggests that convective turbulence may continue to contribute significantly to tidal dissipation even in magnetically influenced regimes. The frequency dependence of the tidal response reveals a clear resonant structure (particularly for large frequency magnitudes) for several magnetic configurations. These peaks become especially clear at larger $\mathrm{Le}$, and are naturally interpreted as Alfvénic (or fast magneto-inertial) mode resonances. To validate this interpretation, we solved the corresponding eigenvalue problem for our system without tidal forcing. We found that the locations of the peaks in the tidal power at high frequency magnitudes agree closely with the least-damped eigenmodes of the linear system. This provides direct evidence that the high-frequency tidal response is governed by resonances between the tidal forcing and weakly damped Alfvénic eigenmodes. 

   We also find that shell thickness plays an important role in controlling magnetic effects. Thinner shells tend to weaken the magnetic influence, with tidal power approaching its hydrodynamical value, especially at moderate $\mathrm{Le}$. This is consistent with enhanced inertial-wave activity in thinner shells. At larger $\mathrm{Le}$, magnetic effects remain present, but high-frequency Alfvénic resonances outside the inertial range become less coherent and decay more rapidly. This indicates that the tidal response is governed by an interplay between magnetic tension and geometric confinement. Within the explored ranges of $\mathrm{Pm}$ and $\mathrm{Ek}$, we find that, in general, the magnetic Prandtl number primarily controls magnetic diffusion rather than the structure of the tidal flow. While tidal power is only weakly affected by $\mathrm{Pm}$, Ohmic dissipation varies significantly, reaching a maximum at intermediate values. Lower $\mathrm{Pm}$ suppresses high-frequency resonances through enhanced diffusion, while leaving the inertial-range response largely unchanged. A similar trend is observed when varying the Ekman number. Decreasing $\mathrm{Ek}$ has little effect on inertial-wave dynamics, but enhances high-frequency Alfvénic resonances by reducing viscous (and Ohmic, for a given Pm) damping. As a result, the overall (frequency-averaged) inertial-wave response is relatively robust, whereas resonances with magnetic waves are more sensitive to dissipative effects.
  
   Despite the strong dependence of the tidal response at individual frequencies on the magnetic field strength and geometry, the frequency-averaged tidal power is found to be much less sensitive to magnetic effects in most cases. For a wide range of parameters and background fields, the numerically computed frequency-averaged tidal power is in good agreement with the prediction derived by \citet{ogilvie2013tides}. This extends the conclusion of previous work \citep{lin2018tidal} to a much broader set of magnetic configurations, and suggests that the long-term secular tidal evolution may remain comparatively insensitive to the detailed magnetic geometry, even though the instantaneous dissipation at a given frequency can vary strongly. However, we do find noticeable discrepancies for sufficiently strong free-decay poloidal dipole and quadrupole fields. These cases likely indicate a breakdown of some of the assumptions underlying the standard frequency-averaged formalism, in particular when the background magnetic field is sufficiently strong, and the work done by Lorentz forces is non-negligible.

   Overall, our results demonstrate that magnetic fields can strongly reshape the detailed spectrum of tidal dissipation in rotating convective regions, especially at individual forcing frequencies and for field configurations with strong poloidal components. Magnetic fields broaden the frequency window for enhanced tidal response, modify the spatial structure of waves, and introduce Alfvénic resonances at high frequencies outside the inertial range. These findings have important implications for modelling tidal evolution in magnetised stars and giant planets, where realistic magnetic fields may substantially alter the dissipation spectrum even if the overall frequency-averaged dissipation is only weakly affected.

\section*{Acknowledgements}
This work was supported by STFC grants ST/W000873/1 and UKRI1179. AA was partly supported by a Leverhulme Trust Early Career Fellowship. We would like to thank the reviewer for a constructive review.

\section*{Data Availability}
The data underlying this article will be shared on reasonable request to the corresponding author.



\bibliographystyle{mnras}
\bibliography{Ref} 
\bsp	
\label{lastpage}
\end{document}